\documentclass[journal,10pt]{IEEEtran}

\usepackage{graphicx,wrapfig,amsmath,float,enumitem,amssymb,verbatim}
\usepackage{setspace,color,soul,subcaption,rotating,pdflscape}
\usepackage[textsize=small, textwidth=1cm]{todonotes}
\usepackage{cite}
\usepackage{nomencl}
\usepackage{color,soul}
\usepackage{esvect}
\makenomenclature

\makeatletter
\def\maketag@@@#1{\hbox{\m@th\normalfont\normalsize#1}}
\makeatother

\usepackage{xcolor}
\usepackage{lineno,hyperref}
\modulolinenumbers[5]
\hypersetup{
    colorlinks  = true, 
    urlcolor    = blue, 
    linkcolor   = blue, 
    citecolor   = black 
}

\begin{document}
\raggedbottom
\title{A Fault-Tolerant Integrated Vehicle Stability Control Using Adaptive Control Allocation}

\author{Ozan Temiz$^1$, Melih Cakmakci$^1$, and Yildiray Yildiz$^1$

\thanks{$^1$Ozan Temiz, Melih Cakmakci (corr. author) and Yildiray Yildiz are with the Department of Mechanical Engineering, Bilkent University, Ankara 06800, Turkey. ozan.temiz.93@gmail.com, [melihc, yyildiz]@bilkent.edu.tr}
}

\maketitle
\IEEEpeerreviewmaketitle
\def \figwidth {3.75in}
\def \todoin#1{\todo[inline]{#1}}

\begin{abstract}
The focus of this paper is an integrated, fault-tolerant vehicle supervisory control algorithm for the overall stability of ground vehicles. Vehicle control systems contain many sensors and actuators that can communicate with each other over communication networks. 
The proposed supervisory control scheme is composed of a high-level controller that creates a virtual control input vector and a low-level control allocator that distributes the virtual control effort among redundant actuators. 
Virtual control input incorporates the required traction force, yaw, pitch, and roll moment corrections, and the lateral force correction to ensure stability while following a maneuvering reference initiated by the driver. 
Based on the virtual control input vector, the allocation module determines front steering angle correction, rear steering angle, traction forces at each tire, and active suspension forces. 
The proposed control framework distinguishes itself from earlier results in the literature by its ability to adapt to failures and uncertainties by updating its parameters online, without the need for fault identification. 
The control structure is validated in the simulation environment using a fourteen degree of freedom nonlinear vehicle model. 
Our results demonstrate that the proposed approach ensures that the vehicle follows references created by the driver despite the loss of actuator effectiveness up to $\mathbf{30}$\% higher longitudinal maneuver velocity and approximately $\mathbf{35}$\% lower roll and pitch angles during steering with representative driving scenarios. 
\end{abstract}

\section{INTRODUCTION}
Over the past two decades, there have been many advancements in the automotive field with the increased cooperation among vehicle subsystems that are traditionally designed separately. 
Vehicle communication networks, low-cost sensors, and dependable mechatronic actuators play an essential role in this new trend, enabling engineers to design an automobile as a single mechatronic system. This new cooperative approach generates redundancies in control problems, as reported in \cite{Cakmakci2010, Ulsoy2012, Dokuyucu2016}. 
Availability of various actuators, together with the cross-coupling between lateral and rotational dynamics, has led to the integrated vehicle control algorithms such as those proposed in \cite{Ataei2018, Mirzaei2017, Fergani2015, Ni2017, Johansen2013, Harkegard2003, yildiz1, yildiz2, Ono2006, Tavasoli2012, Wang2014a}.
These algorithms can exploit coupled dynamics, producing better performance by adjusting actuator inputs accordingly.

Actuation redundancy in ground vehicles can be observed, for example, in yaw rate control (vehicle's rotation about the z-axis). Yaw rate can be regulated using both steering and traction systems:
A vehicle can regulate its yaw rate by altering the amount of longitudinal traction forces at different wheels, thanks to in-wheel electric motors.
Another way of altering the yaw rate is the steering input. While only front-wheel steering was available and the amount of steering was mechanically constrained in the past, steer by wire and four-wheel-steering technology recently became feasible and affordable with the introduction of dependable, low-cost actuators and enabling control strategies \cite{Yin2007} and \cite{Russell2016a}.
Such developments in the industry introduced redundancy and created new opportunities for control researchers.

\begin{figure}[t]
    \begin{center}
        \begin{subfigure}[b]{\linewidth}
            \includegraphics[trim={5cm 12cm 5cm 12cm}, clip=true,width=\linewidth, page=23]{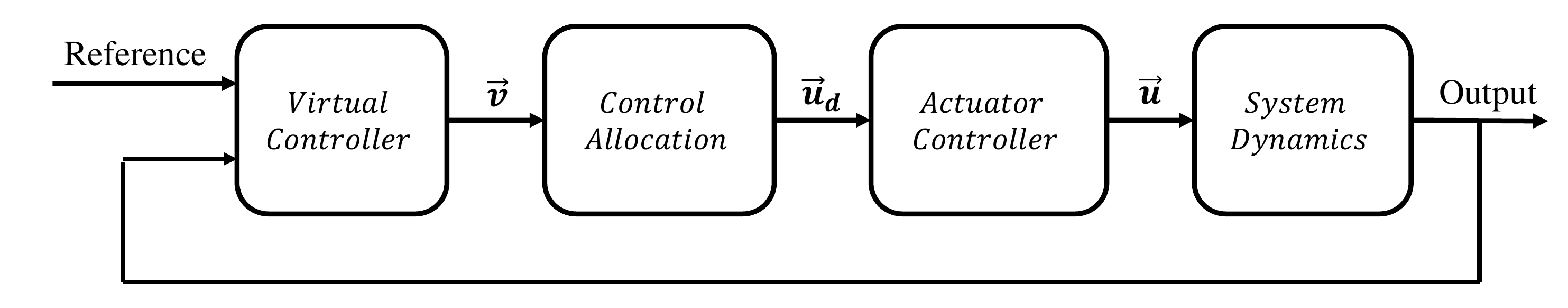}
            \caption[]{{\small}}
        \end{subfigure}
        \begin{subfigure}[b]{\linewidth}
            \includegraphics[trim={5cm 12cm 5cm 12cm}, clip=true,width=\linewidth, page=24]{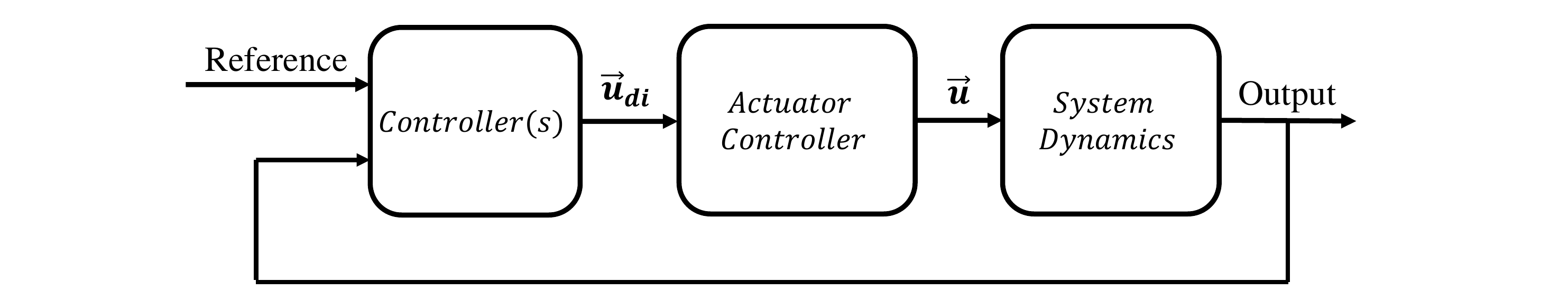}
            \caption[]{{\small}}
        \end{subfigure}
    \end{center}
    \caption{(a) Block diagram of a control system with a control allocator, where $\vec{v}$ is the virtual control input vector, $\vec{u}_d$ is the desired actuator response vector and $\vec{u}$ is the actuator commands vector to achieve the desired response, and (b) block diagram of a control system where the controller produces individual actuator commands, $\vec{u}_{di}$, directly.}
    \label{CA}
\end{figure}

An important challenge to utilize these control redundancies is to find precise approaches that can be implemented for a wide range of vehicle control problems. 
These approaches need to overcome the computational complexity due to the increased number of performance requirements and constraints originating from subsystem controller design problems. 
``Control allocation" \cite{Harkegard2003} can be defined as a systematic way of distributing the total control effort among different actuators.
Control allocation is widely used in flight control to determine the control surface deflections based on a total virtual control input generated by a high-level controller\cite{4124855, Yildiz2010ACA, 5991270}. 
Schemes involving control allocators are usually composed of three successive steps:
In the first step, a virtual control input (for example, the required total force or moment vector to move an object) is determined by a high-level controller such that the overall control objective is met. 
In the second step, the control allocation receives this virtual control input and determines individual actuator commands based on a certain allocation policy.
Finally, as the last step, actuator controllers ensure the realization of these commands. 
Block diagrams of closed-loop control architectures with and without a control allocator are presented in Fig. \ref{CA} as parts (a) and (b), respectively.

There are several novel controller design ideas proposed in the automotive control literature that successfully manage actuator redundancy.
Control allocation for torque vectoring is used in \cite{WONG201622} considering power management, traction control, actuator limits, and fault-cases. 
In \cite{4806119}, a fixed point control allocation algorithm distributes the slip among tires with low computational effort.
In \cite{Wang2016}, a fault-tolerant control architecture is proposed, which can switch between two optimal control allocation schemes. 
In \cite{SHUAI201455}, a control allocation algorithm decides the required torques by taking the controller area network (CAN) communication into consideration.
In \cite{6674064}, an adaptive control allocation is used for energy-efficient path following using regenerative breaking in electric ground vehicles.
In \cite{Tjonnas2010}, another adaptive control allocation application is presented, where the approach given in \cite{Tjonnas2008} is used. 

In this paper, an integrated vehicle controller incorporating an adaptive control allocator is proposed.
In order to ensure that the vehicle follows the driver's intentions, first, a high-level controller generates the virtual control input vector. This vector consists of the desired traction force, yaw, pitch and roll moment, and the required lateral force correction based on the steering and pedal inputs from the driver.
Then, the control allocation algorithm generates the front steering angle correction, rear steering angle, traction forces, and the active suspension forces at each wheel. 
The desired moments and the traction forces are delivered such that the vehicle rotates and accelerates as needed while desired yaw and pitch moments increase the normal forces, when necessary, at individual wheels to provide both stability and driver comfort.
Required lateral force correction ensures that the side-slip angle is in the stable region while following the yaw reference signal.
This paper is a continuation of our work given in \cite{Temiz2018b} and \cite{Temiz2019}. In \cite{Temiz2018b}, a  simpler two-track vehicle model was used for traction control only. In \cite{Temiz2019}, our earlier work was extended with passive suspension dynamics, and the traction control performance in the presence of communication delay in actuators and sensors was studied.

The  contributions of this study can be listed as follows:
(1) A detailed control-oriented nonlinear vehicle model is developed that is suitable for wheel-based traction, steering, and suspension control studies with necessary longitudinal, lateral, and vertical motion fidelity. This model is also validated against a multi-body dynamics based commercial software \cite{Amesim}.
(2)
Inspired from \cite{Tohidi2016}, \cite{TOHIDI20175492} and \cite{tohidinew}, a controller framework involving a fault-tolerant control allocation algorithm is developed, which is capable of controlling the longitudinal, lateral and vertical vehicle dynamics in an integrated manner. To achieve this, the control allocation presented in \cite{Tohidi2016} and \cite{TOHIDI20175492} had to be extended to handle the class of systems that have time-varying control input matrix dynamics. 
To the best of our knowledge, this study is the first of its kind in its attempt to outline a method to develop an integrated vehicle stability controller that utilizes vectoring, steering, and suspension control systems at the same time, in the presence of uncertainties and faults.

The remainder of this paper is structured as follows.
In Section II, the 14 degrees of freedom vehicle model is presented. 
In Section III, the overall structure of the integrated vehicle controller, including the control allocation, is explained. 
Then, the proposed controller is validated via simulation studies in section IV.
Lastly, in Section V, a summary is given.
\section{MATHEMATICAL MODEL}\label{sec:mathematical-model}
\begin{figure}[t]
\begin{center}
\begin{subfigure}[b]{\linewidth}
\includegraphics[trim={17cm 1cm 17cm 0cm},clip=true,height=9cm, width=0.9\linewidth, page=2]{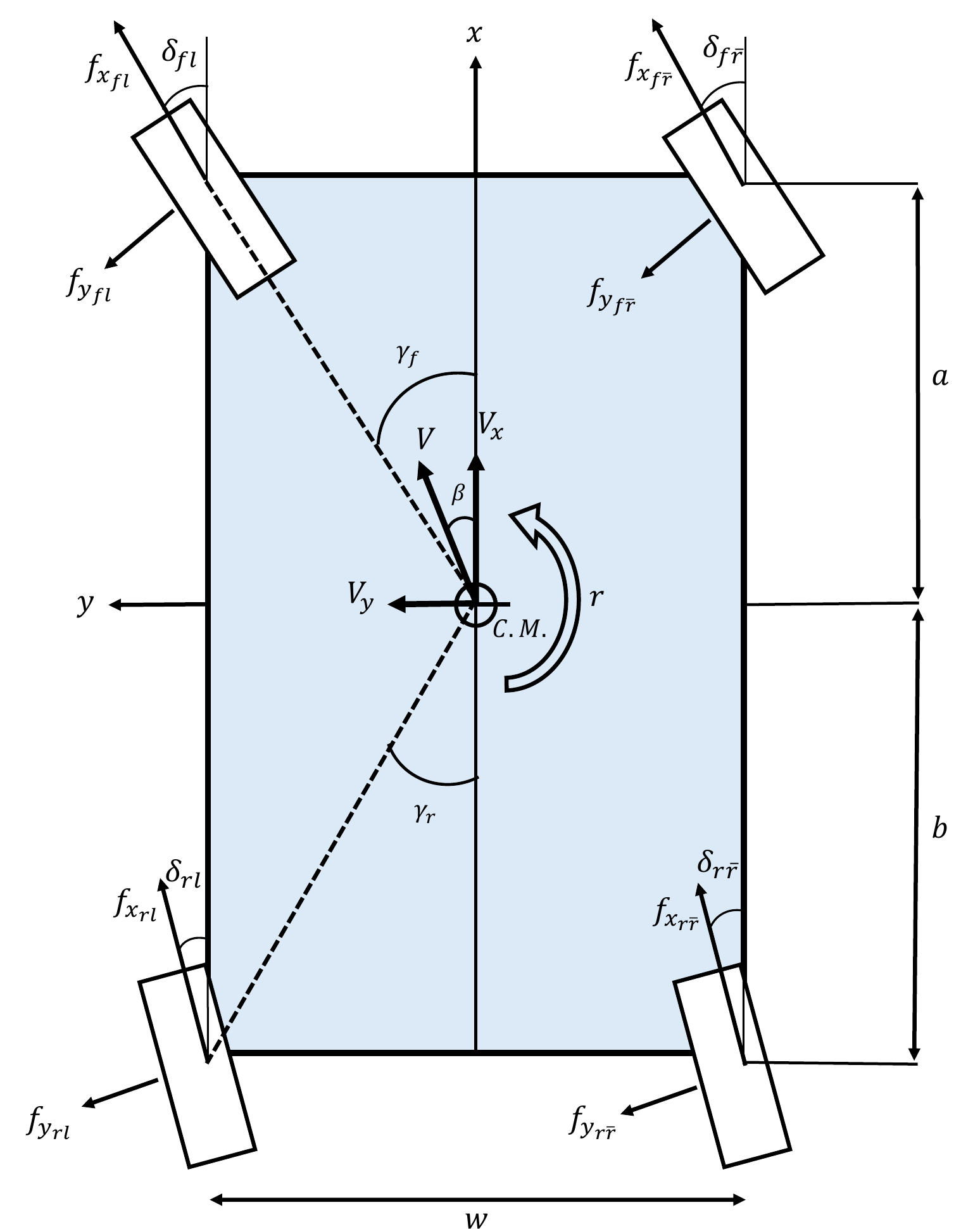}
\caption[]{{\small}Longitudinal and Lateral Dynamics}\label{longandlat} 
\end{subfigure}
\begin{subfigure}[b]{\linewidth}
\includegraphics[trim={1cm 0cm 1cm 0.5cm},clip=true,width=0.9\linewidth, page=3]{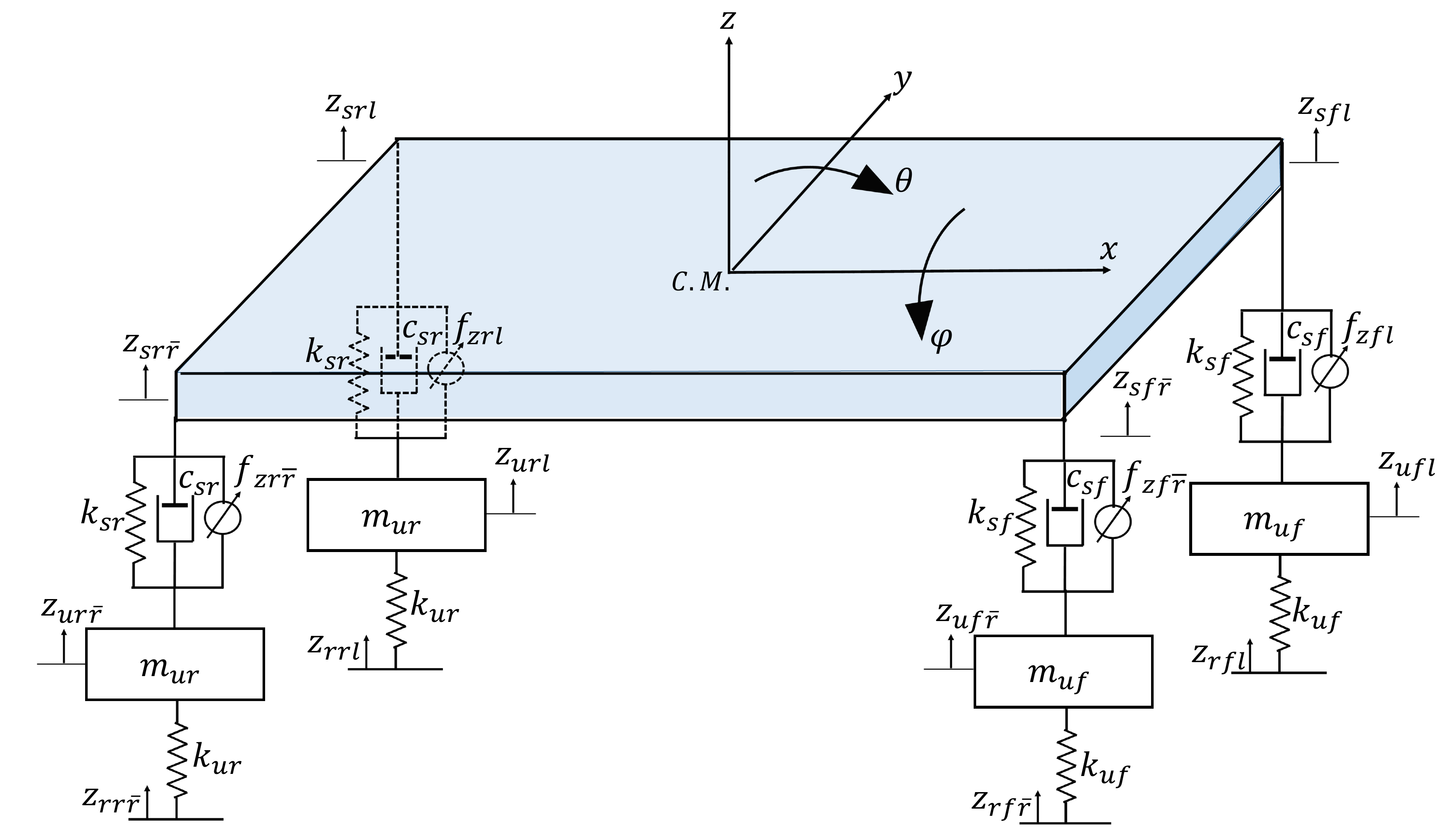}
\caption[]{{\small}Vertical Dynamics}\label{vertical1} 
\end{subfigure}
\end{center}
\caption{Vehicle Dynamic Model}
\label{FBD}
\end{figure}
In this section, the development of a 14 degree of freedom vehicle model is discussed. 
This model includes all three translation and rotation motions of the vehicle body in the three-dimensional space, together with the elevation and the (z-axis) rotation of each wheel.
In order to complete this model, longitudinal, lateral, and vertical motions of the vehicle mass, and each wheel need to be considered (see Fig. \ref{FBD}).
Most models in the literature contain vehicle translational and rotational  motions in one or two directions separately for controller development.
However, the model developed in this section considers the vehicle's motion in three directions, including detailed axis interactions. This model helps to achieve the objective of developing controllers for steering, and suspension systems concurrently. The list of parameters used in the model is given in Table \ref{tab:parameters}.
\begin{table} 
    \captionsetup[table]{skip=5mm}
    \caption{List of Used Vehicle Parameters}    \label{tab:parameters} 
    \centering
    \footnotesize
    \label{table_Journal}
    \begin{tabular}{|c |l| c|}
        \multicolumn{3}{c}{}\\
         \hline
        $h$ & Height of the vehicle Center of Gravity (CoG) & $0.375 \ m$\\
        \hline
        $a$ & Length between the front axle and the CoG & $1.125 \ m$ \\
        \hline
        $b$ & Length between the rear axle and the CoG & $1.375 \ m$\\
        \hline
        $w$ & Width of the wheelbase & $1.6 \ m$\\
        \hline
        $m$ & Vehicle mass & $1300 \ kg$\\
        \hline
        $I_x$ & Veh. moment of inertia about $x-$ axis & $250 \ kg m^2$\\
        \hline
        $I_y$ & Veh. moment of inertia about $y-$ axis & $1000 \ kg m^2$\\
        \hline
        $I_z$ & Veh. moment of inertia about $z-$ axis & $1300 \ kg m^2$\\
        \hline
        $I_w$ & Wheel moment of inertia & $2.7 \ kg m^2$\\
        \hline
        $R_w$ & Wheel radius & $0.33 \ m$\\
        \hline
        $m_{uf}$ & Front unsprung mass & $30 \ kg$\\
        \hline
        $m_{ur}$ & Rear unsprung mass & $30 \ kg$\\
        \hline
        $k_{uf}$ & Front unsprung spring coefficient & $2 \cdot 10^5 \ N/m$\\
        \hline
        $k_{ur}$ & Rear unsprung spring coefficient & $2 \cdot 10^5 \ N/m$\\
        \hline
        $k_{sf}$ & Front sprung spring coefficient & $21 \cdot 10^3 \ N/m$\\
        \hline
        $c_{sf}$ & Front sprung damping coefficient & $1000 \ Ns/m$\\
        \hline
        $k_{sr}$ & Rear sprung spring coefficient & $21 \cdot 10^3 \ N/m$\\
        \hline
        $c_{sr}$ & Rear sprung damping coefficient & $1500 \ Ns/m$\\
        \hline
        $A_f$ & Frontal area of the vehicle& $2.2 \ m^2$\\
        \hline
        $C_d$ & Drag coefficient of the vehicle& $0.3 \ $\\
        \hline
    \end{tabular}
    \normalsize
\end{table} 
\subsection{Longitudinal and Lateral Dynamics}
The dynamic equations describing the longitudinal and lateral motions of the vehicle body can be developed by using the variables and vehicle geometry presented in Fig. \ref{longandlat}. 
Vehicle motion can be described using a coordinate system, $xyz$, which is fixed to the center of mass of the vehicle, as shown in Fig. \ref{vertical1}.

The total forces in the x-direction, $F_x$, and in the y-direction, $F_y$, are calculated as the sum of forces on each tire as
\begin{eqnarray}
F_x =\sum_{i=\{f,r\}}\sum_{j=\{l,\bar{r}\}} F_{xij}
\label{fx}\\
F_y =\sum_{i=\{f,r\}}\sum_{j=\{l,\bar{r}\}} F_{yij},
\label{fy}
\end{eqnarray}
where $f$, $r$, $l$ and $\bar{r}$ indicate front, rear, left and right wheel locations, respectively. 
For example, $F_{xfl}$ represents the longitudinal traction force acting on the front left tire. 
Since each wheel is steerable, these forces can be calculated in terms of the force acting on each tire and the individual steering angles as
\begin{equation}
\begin{aligned}
F_{xfl}&=f_{xfl} \cos\delta_{fl}-f_{yfl} \sin\delta_{fl} \\
F_{yfl}&=f_{yfl} \cos\delta_{fl} + f_{xfl} \sin\delta_{fl} \\
F_{xf\bar{r}}&=f_{xf\bar{r}} \cos\delta_{f\bar{r}}-f_{yf\bar{r}} \sin\delta_{f\bar{r}} \\
F_{yf\bar{r}}&=f_{yf\bar{r}} \cos\delta_{f\bar{r}}+f_{xf\bar{r}} \sin\delta_{f\bar{r}}\\
F_{xrl}&=f_{xrl} \cos\delta_{rl}-f_{yrl} \sin\delta_{rl} \\
F_{yrl}&=f_{yrl} \cos\delta_{rl} + f_{xrl} \sin\delta_{rl} \\
F_{xr\bar{r}}&=f_{xr\bar{r}} \cos\delta_{r\bar{r}}-f_{yr\bar{r}} \sin\delta_{r\bar{r}} \\
F_{yr\bar{r}}&=f_{yr\bar{r}} \cos\delta_{r\bar{r}} + f_{xr\bar{r}} \sin\delta_{r\bar{r}},
\end{aligned}
\label{Forces}
\end{equation}
where $f_{xij}$ and $f_{yij}$ are the longitudinal and the lateral forces, at tire $ij$,  $i \in \{f,r\} $, $j \in \{l,\bar{r}\}$. 
The longitudinal acceleration of the vehicle can be calculated as
\begin{equation}
a_x = \frac{1}{m}[F_x - \frac{1}{2} C_d \rho A_f V_{ar}^2 -mg\sin(Q)],
\label{ax}
\end{equation}
where $C_d$ is the drag coefficient, $\rho$ is the air density, $A_f$ is the frontal area of the vehicle, $V_{ar}$ is the relative velocity between the displaced air and the vehicle, $Q$ is the road slope, and $m$ is the mass of the vehicle. 
Similarly, lateral acceleration, $a_y$, can be calculated as
\begin{equation}
a_y =\frac{1}{m}F_y.
\label{ay}
\end{equation}
It is important to note that the coordinate system $[x,y,z]$ moves and rotates with the vehicle.
Therefore, the acceleration calculations in \eqref{ax} and \eqref{ay} are given for an inertial coordinate system that coincides with the $x$ and $y$ directions at the moment.
The side-slip angle, $\beta$, of the vehicle motion is defined as
\begin{equation}
\beta = \arctan \left(\frac{V_x}{V_y} \right),
\end{equation}
where $V_x$ and $V_y$ are the velocity components of the vehicle measured in the $[x,y,z]$ coordinate system. 
Furthermore, the derivative of the yaw rate, $\dot{r}$, can be calculated as
\begin{equation}
\begin{aligned}
\dot{r} &= \frac{1}{I_z}(\frac{w}{2} (F_{xf\bar{r}}+F_{xr\bar{r}}-F_{xfl}-F_{xrl}) \\
& +(F_{yfl}+F_{yf\bar{r}})a - (F_{yrl}+F_{yr\bar{r}})b),
\label{yaw}
\end{aligned}
\end{equation}
where $I_z$ is the moment inertia about the $z$ axis, $w$ is the width of the wheelbase, and $a$, $b$ are the distances between the center of gravity and front and rear wheels, respectively (see Fig. \ref{longandlat}).
\begin{figure}[t]
\begin{center}
\includegraphics[trim={1cm 0cm 1cm 0cm},clip=true,width=\linewidth, page=4]{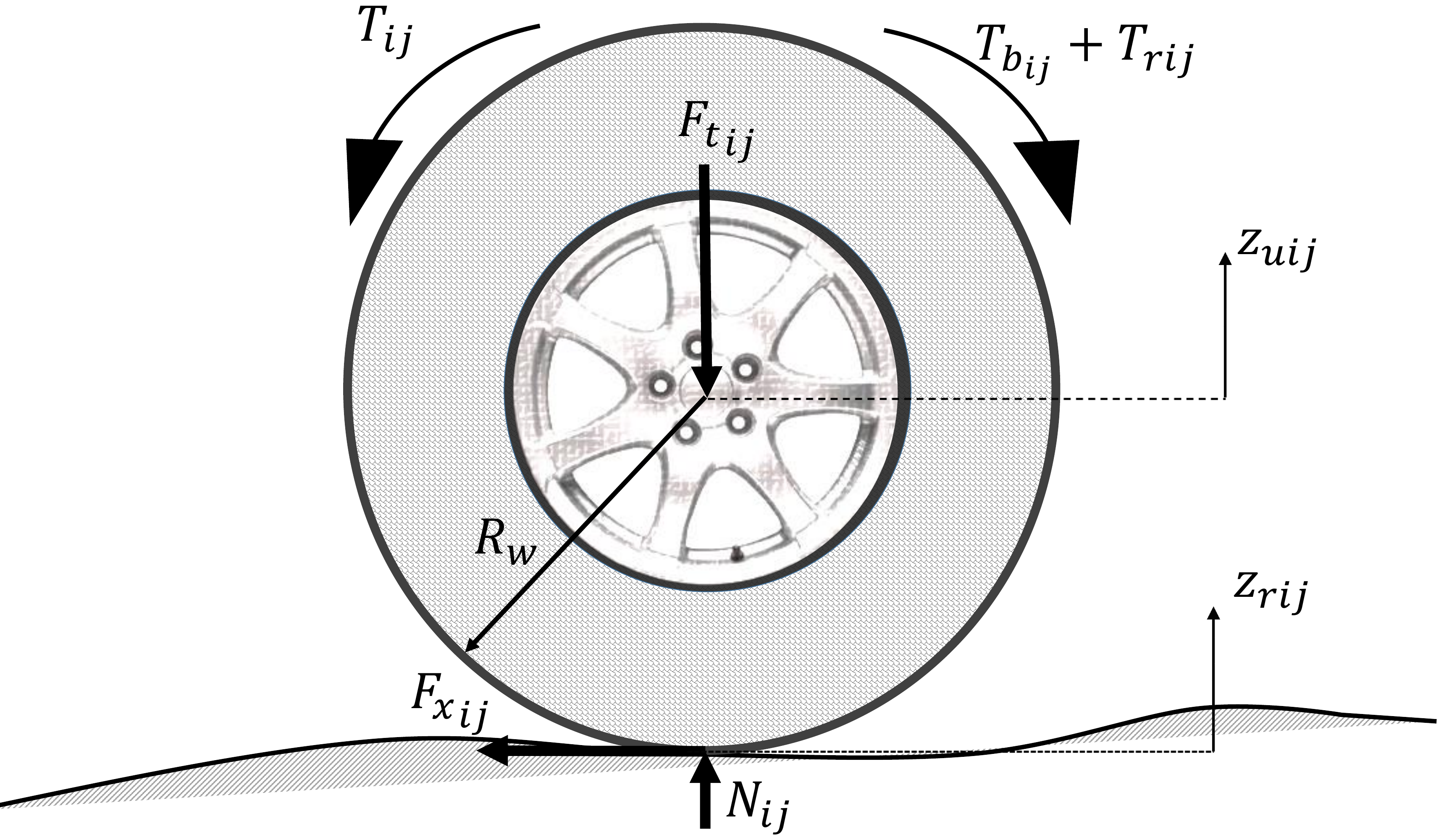}
\end{center}
\caption{Tire Dynamics}
\label{wheel} 
\end{figure}

Wheel assembly, including the tire, is responsible for creating the traction forces for vehicle acceleration or deceleration. 
As the wheel spins with the torque input, the rotation starts, and the resulting slip between tire and road causes the traction force.
Wheel dynamics are typically represented with the rotation of the tire about its lateral axis (traction) and the vertical axis (steering). 
For this work, it is assumed that the driver steering input is realized without any transient dynamics. 
The rotational dynamics equation about the lateral axis, for each wheel, can be obtained as
\begin{equation}
I_{w} \dot{w}_{ij}= T_{ij}-T_{bij}-T_{rij} -F_{xij}R_w,
\label{tire}
\end{equation}
where, $\omega_{ij}$, $T_{rij}$ $T_{ij}$, and $T_{bij}$, are the angular velocity, rolling resistance, motor, and brake torques, respectively, that are applied on the wheel $ij$, $i \in \{f,r\} $, $j \in \{l,\bar{r}\}$. 
\nomenclature{$\omega_{ij}$}{Rotational velocity of the wheel <\{f,r\}><\{l,r\}>}
The rolling resistance, which originates from the deformable nature of the tire \cite{ Genta2009} is calculated as
\begin{equation}
T_{rij}= p_0 N_{ij} + p_1 N_{ij} \frac{V_x}{30} + p_2 N_{ij} \frac{V_x^4}{30^4},
\label{rolling_res}
\end{equation}
where, $N_{ij}$ is the normal force acting on tire $ij$. The coefficients $p_0$, $p_1$ and $p_2$ are usually obtained experimentally and typical values can be found in \cite{Kiencke2005}.
In order to calculate the longitudinal and lateral forces, $f_{xij}$ and $f_{yij}$, respectively, on the tire,  a  friction model is required. In this work, Pacejka's Magic tire formula \cite{Pacejka2006} is used. 
According to this formula, longitudinal traction force $f_{xij}$ can be calculated as
\begin{equation}
\begin{aligned}
f_{xij}=& D_1 \sin[C_1 \arctan (B_1 \lambda_{ij}-E_1(B_1 \lambda_{ij} \\
& - \arctan(B_1 \lambda_{ij}) ))],
\label{magic1}
\end{aligned}
\end{equation}
where $D_1$ is the peak value, $C_1$ is the shape factor, $B_1$ is the stiffness factor and $E_1$ is the curvature factor.
The longitudinal slip, $\lambda_{ij}$, in \eqref{magic1} is calculated separately for driving and braking conditions as
\begin{equation}
\begin{aligned}
For \  driving; \quad \lambda_{ij}(V_x,w_{ij})&=\frac{w_{ij} R_w-V_x}{w_{ij} R_w} \\
For \  braking; \quad \lambda_{ij}(V_x,w_{ij})&=\frac{w_{ij} R_w-V_x}{V_x},
\end{aligned}
\label{long_slip}
\end{equation}
where $V_x$ is the vehicle longitudinal velocity, $w_{ij}$ is the rotational velocity on wheel $ij$, $i \in \{f,r\} $, $j \in \{l,\bar{r}\}$, and $R_w$ is the radius of the wheel. 
The slip ratios, $\alpha_{ij}(V_x, V_y, r) $, for each tire, can be calculated using vehicle geometry as
\begin{equation}
\begin{aligned}
\alpha_{fl}(V_x,V_y,r) &= \delta_{fl}- \arctan \left( \frac{V_y + r \ a \ cos(\gamma_f)}{V_x - r \ a \ \sin(\gamma_f)} \right) \\
\alpha_{f\bar{r}}(V_x,V_y,r) &= \delta_{fr}- \arctan \left( \frac{V_y + r \ a \ cos(\gamma_f)}{V_x + r \ a \ \sin(\gamma_f)} \right) \\
\alpha_{rl}(V_x,V_y,r) &= \delta_{rl}- \arctan \left( \frac{V_y -r \ b \ cos(\gamma_r)}{V_x - r \ b \ \sin(\gamma_r)} \right) \\
\alpha_{r\bar{r}}(V_x,V_y,r) &= \delta_{rr}- \arctan \left( \frac{V_y -r \ b \ cos(\gamma_r)}{V_x + r \ b \ \sin(\gamma_r)} \right).
\end{aligned}
\label{lateral_frictions}
\end{equation}
In \eqref{lateral_frictions}, $\delta_{ij}$ is the steering angle on wheel $ij$, $i \in \{f,r\} $, $j \in \{l,\bar{r}\}$. 
The front and rear hub angles, $\gamma_f$ and $\gamma_r$, are shown in Fig. \ref{FBD}. 
Finally, the lateral forces, $f_{yij}$, can be calculated by applying the Magic Formula given \cite{Pacejka2006} as
\begin{equation}
\begin{aligned}
f_{yij} =& D_2 \sin[C_2 \arctan (B_2 \alpha_{ij} -E_2(B_2 \alpha_{ij} \\
& - \arctan(B_2 \alpha_{ij})))].
\label{magic2}
\end{aligned}
\end{equation}

\subsection{Vertical Dynamics}
One of the main challenges of designing a comfortable car is finding suspension parameters that provide a balance between comfort and handling. 
A vertical model integrated with the handling model is needed to consider these two performance requirements simultaneously (see Fig. \ref{FBD}). 
The vertical dynamics model in this study is developed in two parts. 
First, equations of motion for sprung mass, which is the vehicle body supported by springs, are written. 
Then, unsprung masses, which are the wheels that are excited from road disturbances, are modeled. 
It is assumed that the suspension system has a linear spring and a damper, while the tire is assumed to act as a linear spring \cite{Ikenaga2000}. 

Using the pitch, $\theta$, and  the roll, $\phi$, angles  the sprung mass elevation at each wheel location, $z_{sij}$, can be calculated as 
\begin{eqnarray}
z_{sfl}=z-a \sin(\theta)+0.5 w \sin(\phi), \\
z_{sf\bar{r}}=z-a \sin(\theta)-0.5 w \sin(\phi),\\
z_{srl}=z+b \sin(\theta)+0.5 w \sin(\phi), \\
z_{sf\bar{r}}=z+b \sin(\theta)-0.5 w \sin(\phi),
\end{eqnarray}
where $z$ is the elevation of the center of mass (heave). 

Using Fig. \ref{FBD}, the heave acceleration ($\ddot{z}$) can be calculated as
\begin{equation}
\begin{aligned}
\ddot{z}&= \frac{1}{m}((2k_{sf}+2k_{sr})z-(2c_{sf}+2c_{sr})\dot{z}\\&+(2ak_{sf}-2bk_{sr})\sin(\theta)+(2ac_{sf}-2bc_{sr})\dot{\theta}\cos(\theta)\\&+k_{sf}z_{ufl}+c_{sf}\dot{z}_{ufl} +k_{sf}z_{uf\bar{r}}+c_{sf}\dot{z}_{uf\bar{r}}\\&+k_{sr}z_{url}+c_{sr}\dot{z}_{url}+k_{sr}z_{ur\bar{r}}+c_{sr}\dot{z}_{ur\bar{r}} \\& + f_{zf\bar{r}} + f_{zfl} + f_{zr\bar{r}} + f_{zfl}),
\end{aligned}
\label{heave}
\end{equation}
where $k_{sf}$ and $c_{sf}$ are the front sprung mass spring and damping coefficients, $k_{sr}$ and $c_{sr}$ are the rear sprung mass spring and damping coefficients, respectively. $z_{uij}$ is the unsprung mass displacement and $f_{ij}$ is the active suspension forces at wheel $ij$, $i \in \{f,r\} $, $j \in \{l,\bar{r}\}$.
Using Fig. \ref{vertical1}, the pitch acceleration $\ddot{\theta}$ can be calculated as
\begin{equation}
\begin{aligned}
\ddot{\theta}&= \frac{1}{I_y}((2ak_{sf}-2bk_{sr})z+(2ac_{sf}-2bc_{sr})\dot{z}\\&-(2a^2k_{sf}+2b^2k_{sr})\sin(\theta)-(2a^2c_{sf}+2b^2c_{sr})\dot{\theta}\cos(\theta)\\&-ak_{sf}z_{ufl} -ac_{sf}\dot{z}_{ufl}-ak_{sf}z_{uf\bar{r}}-ac_{s}\dot{z}_{uf\bar{r}}\\&+bk_{sr}z_{url}+bc_{sr}\dot{z}_{url}+bk_{sr}z_{ur\bar{r}}+bc_{sr}\dot{z}_{ur\bar{r}}\\& - ma_x h - a f_{zfl} - a f_{zf\bar{r}} + b f_{zrl} + b f_{zr\bar{r}}).
\end{aligned}
\label{pitch1}
\end{equation}
Similarly, the roll  acceleration $\ddot{\phi}$ can be calculated as
\begin{equation}
\begin{aligned}
\ddot{\phi}&=\frac{1}{I_x}(-0.25w^2(2k_{sf}+2k_{sr})\sin(\phi)+0.5wk_{sf}z_{ufl}\\&-0.25w^2(2c_{sf}+2c_{sr})\dot{\phi}\cos(\phi)+0.5wc_{sf}\dot{z}_{ufl}\\&-0.5wk_{sf}z_{uf\bar{r}} -0.5wc_{sf}\dot{z}_{uf\bar{r}}+0.5wk_{sr}z_{url}\\&+0.5wc_{sr}\dot{z}_{url}-0.5wk_{sr}z_{ur\bar{r}} -0.5wc_{sr}\dot{z}_{ur\bar{r}}\\& - ma_y h + \frac{w}{2} f_{zfl} - \frac{w}{2} f_{zf\bar{r}} + \frac{w}{2} f_{zrl} - \frac{w}{2} f_{zr\bar{r}}).
\end{aligned}
\label{roll}
\end{equation}
%

Elevation of each unsprung mass, $z_{uij}$, $i \in \{f,r\}$, $j \in \{l,\bar{r}\}$, can be calculated by using the forces acting on the wheels as
\begin{equation}
\begin{aligned}
m_{uf} \ddot{z}_{ufl}&= k_{sf}z+c_{sf}\dot{z}-ak_{sf}\sin(\theta)-ac_{sf}\dot{\theta}\cos(\theta)\\&+0.5wk_{sf}\sin(\phi)+0.5wc_{sf}\dot{\phi}\cos(\phi)\\&-(k_{sf}+k_{uf})z_{ufl}-c_{sf}\dot{z}_{ufl}\\&+k_{uf}z_{rfl} -f_{zfl},
\end{aligned}
\label{vertical}
\end{equation}
\begin{equation}
\begin{aligned}
m_{uf} \ddot{z}_{ufr}&= k_{sf}z+c_{sf}\dot{z}-ak_{sf}\sin(\theta)-ac_{sf}\dot{\theta}\cos(\theta)\\&-0.5wk_{sf}\sin(\phi)-0.5wc_{sf}\dot{\phi}\cos(\phi)\\&-(k_{sf}+k_{uf})z_{ufl}-c_{sf}\dot{z}_{ufr}\\&+k_{uf}z_{rfr} -f_{zf\bar{r}},
\end{aligned}
\end{equation}
\begin{equation}
\begin{aligned}
m_{ur} \ddot{z}_{url}&= k_{sr}z+c_{sr}\dot{z}+ak_{sr}\sin(\theta)+ac_{sr}\dot{\theta}\cos(\theta)\\&+0.5wk_{sr}\sin(\phi)+0.5wc_{sr}\dot{\phi}\cos(\phi)\\&-(k_{sr}+k_{ur})z_{url}-c_{sr}\dot{z}_{url}\\&+k_{ur}z_{rrl} -f_{zrl},
\end{aligned}
\end{equation}
and
\begin{equation}
\begin{aligned}
m_{ur} \ddot{z}_{urr}&= k_{sr}z+c_{sr}\dot{z}+ak_{sr}\sin(\theta)+ac_{sr}\dot{\theta}\cos(\theta)\\&-0.5wk_{sr}\sin(\phi)-0.5wc_{sr}\dot{\phi}\cos(\phi)\\&-(k_{sr}+k_{ur})z_{urr}-c_{sr}\dot{z}_{urr}\\&+k_{ur}z_{rrr} -f_{zr\bar{r}},
\end{aligned}
\end{equation}
where, $z_{rij}$ is the road disturbance at wheel $ij$, $m_{uf}$, $m_{ur}$, $k_{uf}$ and $k_{ur}$ are the sprung masses and the unsprung spring coefficients at the front and the rear tires, respectively. 
Finally, the normal forces at each tire can be calculated as
\begin{equation}
N_{ij} = k_{si}(z_{uij} - z_{rij}),
\label{normalforces}
\end{equation}
where $N_{ij}$ is the normal force at tire $ij$.

\begin{figure*}[t]
    \begin{center}
        \includegraphics[page=1,trim={3cm 3cm 5cm 3cm},clip=true,width=\linewidth]{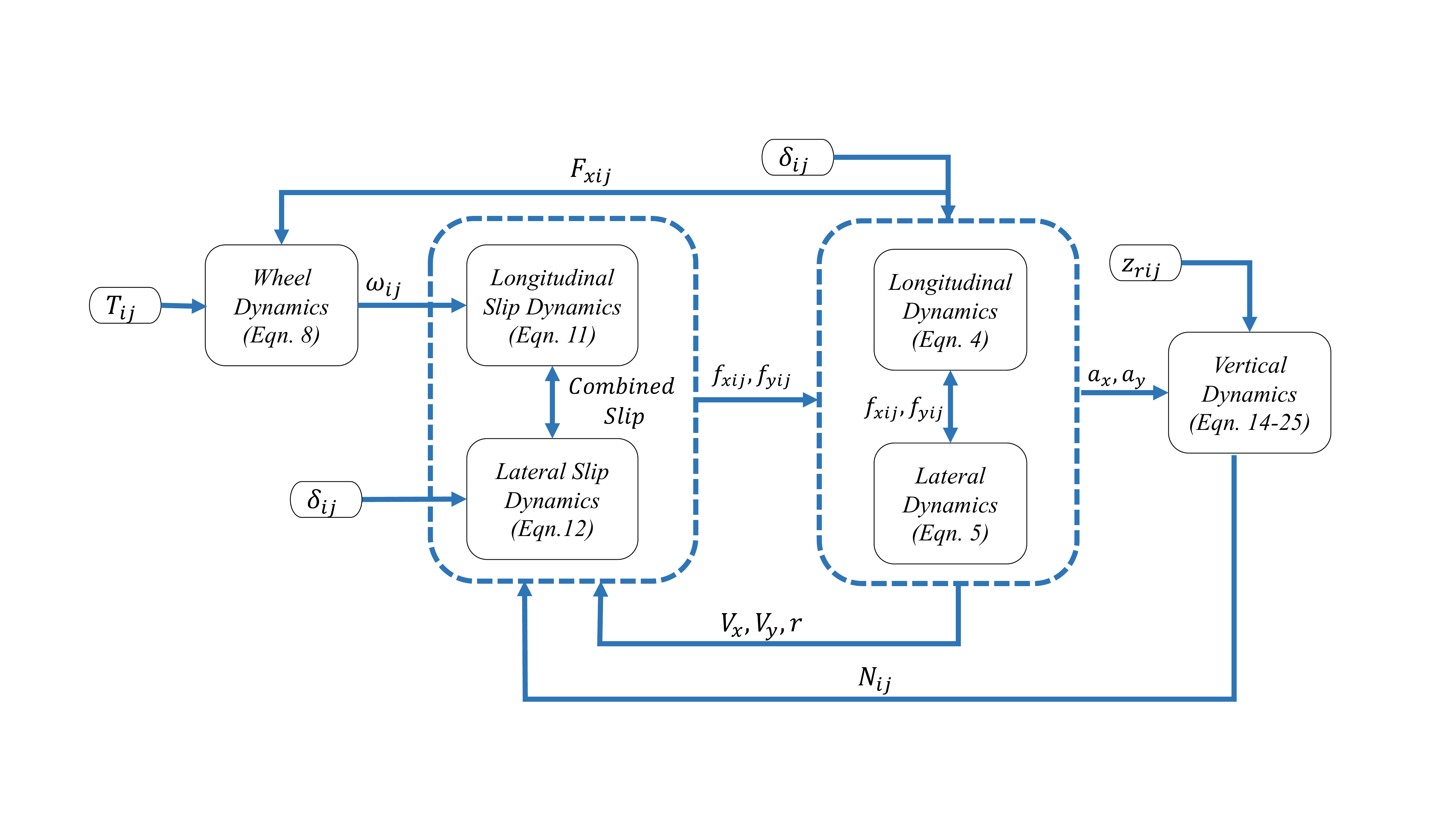}
    \end{center}
    \caption{Vehicle Model and Its Components}
    \label{fig:model_structure} 
\end{figure*}

Fig. \ref{fig:model_structure} summarizes the interaction among many components of the resulting vehicle model developed in \eqref{fx}-\eqref{normalforces}.
The inputs of the model are the wheel torque, $T_{ij}$, the steering angle, $\delta_{ij}$ and the road excitation, $z_{rij}$, at each wheel, and the outputs are the all six translational and rotational motions of the vehicle as well as the rotation of the wheels and the deflections at the wheels and suspensions. 

The proposed model presented in \eqref{fx}-\eqref{normalforces} is validated against the multi-body dynamics based commercial software \cite{Amesim} and found in good correlation for the driving conditions used in this work. 
Results for the validation studies can be found in \cite{otemizthesis}.
\section{CONTROL STRUCTURE}\label{sec:control-structure}
\begin{figure*}[t]
    \begin{center}
        \includegraphics[page=5,trim={0.1cm 0.3cm 0.1cm 0.3cm},clip=true,width=\linewidth]{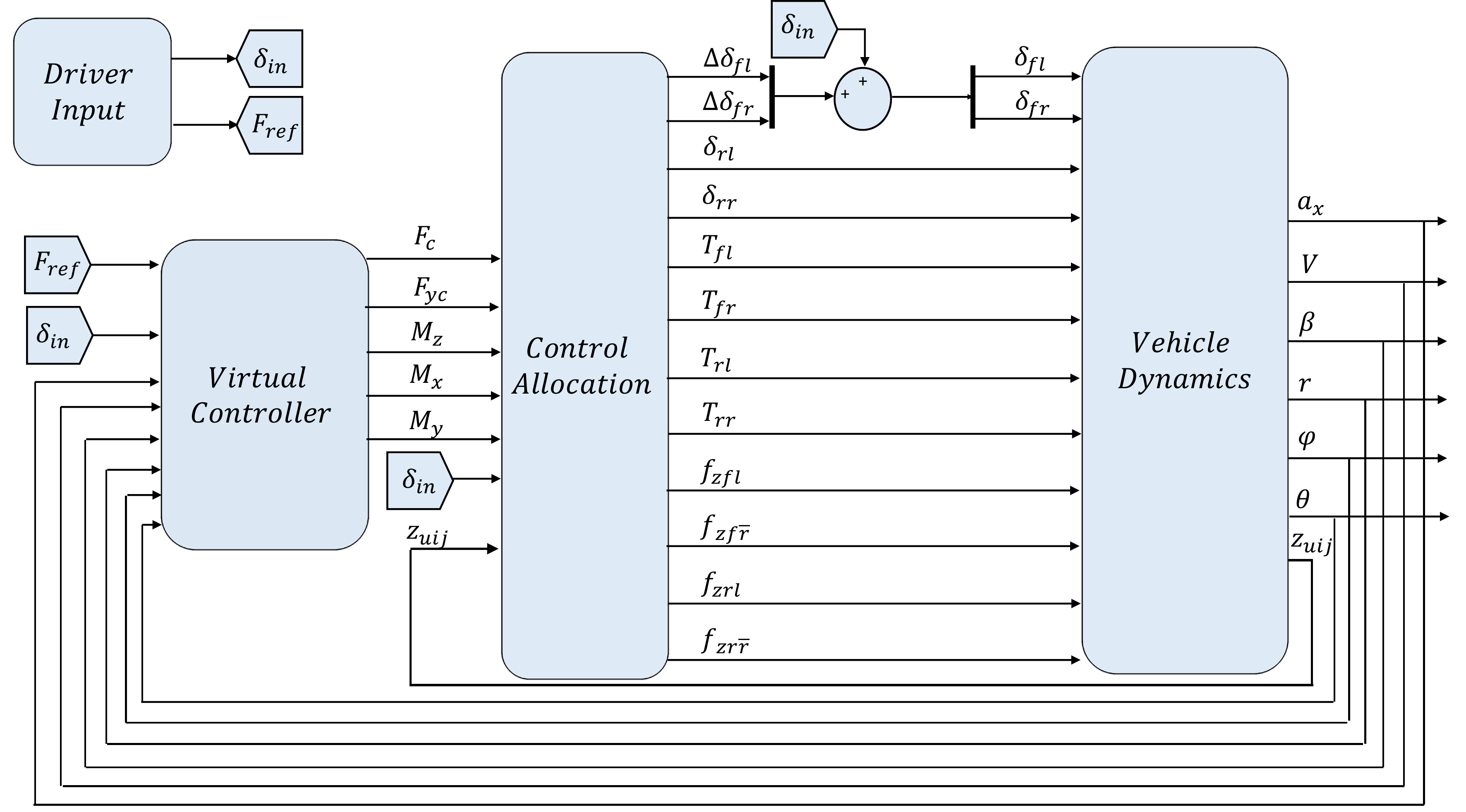}
    \end{center}
    \caption{Overall Closed Loop System}
    \label{control_structure} 
\end{figure*}
The overall closed-loop system, including the proposed control structure, is given in Fig. \ref{control_structure}. 
In this structure, the driver provides the steering and the desired traction torque (based on the pedal position) inputs. 
These inputs are used by the controller to produce the virtual control input vector, containing the desired traction force, $F_c$, and the desired pitch, roll, and yaw moment corrections, namely $M_{x}$, $M_{y}$ and $M_{z}$, respectively.
Moreover, to ensure yaw stability, the desired lateral force correction, $F_{yc}$, is also calculated as another component of the virtual control input vector. 
The proposed control allocation algorithm determines the torque $T_{ij}$ to be applied at each wheel, the rear-wheel steering angles $\delta_{rl}$, $\delta_{r\bar{r}}$, the front-wheel steering angle corrections $\Delta\delta_{fl}$,  $\Delta\delta_{f\bar{r}}$, and the active suspension forces $f_{ij}$, $i \in \{f,r\}$, $j \in \{l,\bar{r}\}$, based on the virtual control input vector. 
\subsection{Virtual Control Input Generation}\label{sec:virtual-control-input-generation}

In this section, calculation of the virtual control input vector elements, which are the desired traction force $F_c$, desired yaw, roll and pitch moment corrections $M_z$, $M_{x}$, $M_{y}$, and the required lateral force correction, $F_{yc}$, is described.
The traction force command, $F_c$, is calculated using a Proportional Integral (PI) controller to ensure that the vehicle maintains the desired longitudinal acceleration:
\begin{equation}
\begin{gathered}
\tilde{F}=F_{ref}-F \\
F_c=K_{if} \int{\tilde{F}} dt +K_{pf}\tilde{F},
\end{gathered}
\label{fc2}
\end{equation}
where $F_{ref}$ is the traction force mapped from the pedal input of the driver,  $F$ is the vehicle traction force, and $K_p$ and $K_i$ are the PI controller gains. 

To follow the steering input of the driver, first, a  reference yaw rate, $r_{ref}$, is calculated using the methods proposed in \cite{rajamani2011vehicle} and \cite{Ulsoy2012}. 
Then, by defining the error, $\tilde{r} = r_{ref} - r$, between $r_{ref}$ and the measured yaw rate, $r$, a moment, $M_1$, is generated as
\begin{equation}
M_1 = K_{pmz} \tilde{r}+K_{imz} \int\tilde{r}dt,
\end{equation}
where $K_{pmz}$ and $K_{imz}$ are the PI controller gains. 
Additionally, to ensure stability while following the yaw rate reference, side-slip angle $\beta$ is fed to another PI controller to produce an additional moment, $M_2$, as
\begin{equation}
M_2 = K_{ps} \beta + K_{is} \int\beta dt,
\end{equation}
where, $K_{ps}$, $K_{is}$ are the PI controller gains. 
Moments $M_1$ and $M_2$ are then summed to create the desired yaw moment correction $M_z$.
\begin{equation}
\begin{gathered}
M_z= M_1 + M_2.
\label{Mz}
\end{gathered}
\end{equation}
Lateral and longitudinal accelerations may cause the vehicle to roll and pitch. 
These motions are not desirable since they shift the center of gravity and may disturb the vehicle's stability. 
Therefore, to damp the roll and pitch motions, roll and pitch moment corrections, $M_x$ and $M_y$, are calculated by using a PI controller as
\begin{equation}
\begin{gathered}
M_{x}=- K_{pr} \phi - K_{dr} \dot{\phi} - K_{ir} \int\phi dt \\
M_{y}=- K_{pp} \theta - K_{dp} \dot{\theta} - K_{ip} \int\theta dt,
\end{gathered}
\label{roll_pitch}
\end{equation}
where $\phi$ and $\theta$ are the roll and pitch angles of the vehicle and $K_{pr}$, $K_{dr}$, $K_{ir}$, $K_{pp}$, $K_{dp}$ and $K_{ip}$ are the PID controller gains.

The side-slip angle, $\beta$, should be kept small since a large value causes the vehicle to be unstable \cite{Tjonnas2010}. 
$\beta$ can be controlled by applying a lateral force $F_{yc}$, which can be calculated as
\begin{equation}
F_{yc}= - K_{py} \beta - K_{iy} \int{\beta},
\label{Fy_law}
\end{equation}
where $K_{py}$ and $K_{iy}$ are the PI controller constants. 

\subsection{Vehicle Model for Control Allocation}

In order to develop a control allocator, a control allocation oriented model is derived from the nonlinear vehicle dynamics model presented in Section \ref{sec:mathematical-model}. 

The vehicle outputs, which are used in the feedback loop (see Fig. \ref{control_structure}), are the vehicle acceleration, $a_x$, the vehicle velocity, $V$, yaw rate, $r$, roll angle, $\phi$, pitch angle, $\theta$, and the wheel hub elevations, $z_{uij}$, $i \in \{f,r\} $, $j \in \{l,\bar{r}\}$. 
The states of the control-oriented model, which are the minimum number of variables the initial conditions of which are required to be known to predict the future behavior of the vehicle outputs, are then determined as
\begin{equation}
 \begin{aligned}
 \mathbf{x}^T=
\begin{split}
&[\begin{matrix}V_x & V_y & \dot{\psi} & z & \dot{z} & \phi & \dot{\phi} & \theta & \dot{\theta} & z_{ufl} \end{matrix} \\
&\begin{matrix} \dot{z}_{ufl} & z_{uf\bar{r}} & \dot{z}_{uf\bar{r}} & z_{url} & \dot{z}_{url} &  z_{ur\bar{r}} & \dot{z}_{ur\bar{r}} \end{matrix}].
\end{split}
\end{aligned}
 \label{Xmatrix}
\end{equation}
Using these states, non-linear dynamics of the vehicle given in \eqref{fx}-\eqref{normalforces} can be represented in the form,
\begin{equation}
\begin{aligned}
\dot{\mathbf{x}} = f(\mathbf{x},\mathbf{u}),
\end{aligned}
\label{model}
\end{equation}
where the state vector, $\mathbf{x}$, is given in \eqref{Xmatrix}, and the actuator input vector, $\mathbf{u}$, is
\begin{equation}
\begin{aligned}
\mathbf{u}^T &= 
\begin{split}
&[\begin{matrix}
\delta_{fl} & \delta_{f\bar{r}} & \delta_{rl} & \delta_{r\bar{r}} & T_{fl} & T_{f\bar{r}} \end{matrix} \\ &\begin{matrix} T_{rl} & T_{r\bar{r}} & f_{zfl} & f_{zf\bar{r}} & f_{zrl} & f_{zr\bar{r}}]. \
\end{matrix}
\end{split}
\end{aligned}
\label{umatrix}
\end{equation}
Assuming a constant cruising velocity $V_{0}$, and small steering ($\delta_{ij}$) and tire slip angles  ($\alpha_{ij}$), a linearized time-varying vehicle model can be obtained in the form
\begin{equation}
\dot{\mathbf{x}}=\mathbf{A} \mathbf{x} + \mathbf{B_u}(t) \mathbf{u} + \mathbf{D},
\label{lin_model}
\end{equation}
where $\mathbf{A}$ $\in R^{17 \times 17}$ is the state matrix, $\mathbf{B_u}(t)$ $\in R^{17 \times 12}$ is the time varying input matrix and $\mathbf{D} $ $\in R^{17}$ contains the disturbances. 
The contents of the matrices $\mathbf{A}$, $\mathbf{B_u}(t)$, and $\mathbf{D}$ are available online \footnote{\url{https://github.com/otemiz/Adaptive_control_allocation}}.

To treat the elements of u given in \eqref{umatrix} as pure force and moment generators, and hence make the representation suitable for control allocation, rows of the input matrix $B_u$ that corresponds to the variations in heave, $z$, and unsprung mass, $z_{uij}$, elevations are taken to be zero. 
This assumption ignores the effects of active suspension forces on the vehicle body and unsprung mass accelerations. This is a common practice for control allocation implementations, the examples of which can be seen at \cite{Harkegard2003, Harkegard2005137, Harkegard2004}. 
It is noted that the linear time-varying dynamics \eqref{lin_model} is used only for control allocation development purposes. For the controller validation tests, the full nonlinear model developed in Section \ref{sec:mathematical-model} is employed. 

\subsection{Control Allocation}

\begin{figure}
    \begin{center}
        \includegraphics[page=6,trim={3cm 9.25cm 3cm 9.5cm},clip=true,width=\linewidth]{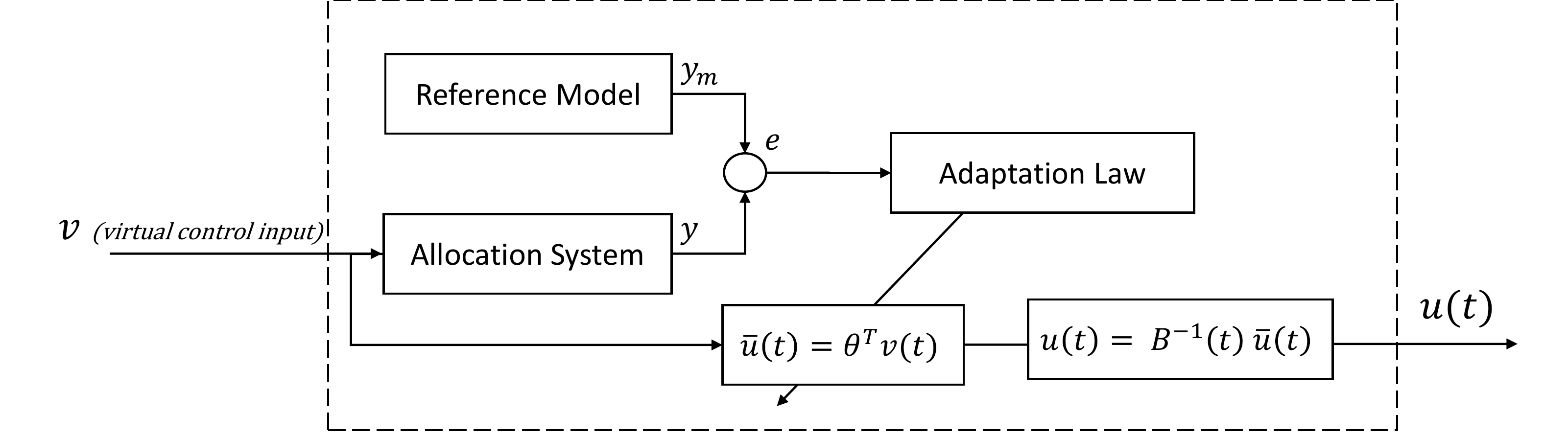}
    \end{center}
    \caption{Control Allocation Structure}
    \label{allocation} 
\end{figure}

The exploited control allocation method is based on \cite{Tohidi2016}, and the block diagram for this algorithm, modified for the specific application considered in this paper, is given in Fig. \ref{allocation}.
For the vehicle control application studied in this paper, the algorithm is re-worked such that it can handle the time-varying input matrix $\mathbf{B_u}(t)$ introduced in \eqref{lin_model}. 

Rewriting $\mathbf{u}$ as a summation of the driver input, $\mathbf{\Delta u}$, and the control allocation input, $\mathbf{u}_{ca}$,  (i.e. $\mathbf{u}  \equiv  \mathbf{u}_{ca}+\mathbf{\Delta u}$), \eqref{lin_model} can be rewritten as
\begin{equation}
\begin{aligned}
\dot{\mathbf{x}}&=\mathbf{A} \mathbf{x} + \mathbf{B_u}(t) \mathbf{u}_{ca} + \mathbf{B_u}(t) \mathbf{\Delta u} + \mathbf{D},
\end{aligned}
\label{plant1}
\end{equation}
where 
\begin{equation}
\begin{aligned}
\mathbf{\Delta u}^T &= 
\begin{split}
&[\begin{matrix} \delta_{in} & \delta_{in} & 0 & 0 & 0 & 0 \end{matrix} \\
&\begin{matrix} 0 & 0 & 0 & 0 & 0 & 0 \end{matrix}],
\end{split}
\\ \mathbf{u}_{ca}^T &= 
\begin{split}
&[\begin{matrix}
\Delta\delta_{fl} & \Delta\delta_{f\bar{r}} & \delta_{rl} & \delta_{r\bar{r}} & T_{fl} & T_{f\bar{r}} \end{matrix} \\
&\begin{matrix} T_{rl} & T_{r\bar{r}} & f_{zfl} & f_{zf\bar{r}} & f_{zrl} & f_{zr\bar{r}} \end{matrix}].
\end{split}
\end{aligned}
\label{plant1_2}
\end{equation}
In accordance with this decomposition, $\mathbf{B_{u}}(t)$ can be split into matrices  $\mathbf{B_v} \in R^{17 \times 5}$ and $\mathbf{B_y}(t) \in R^{5 \times 12}$, as $\mathbf{B_u}(t) = \mathbf{B_v} \mathbf{B_y}(t)$. Matrices $\mathbf{B_v}$ and $\mathbf{B_y}(t)$ are given in Appendix A. 
Using this decomposition, \eqref{plant1} can be rewritten as
\begin{equation}
\dot{\mathbf{x}}=\mathbf{A} \mathbf{x} + \mathbf{B_v} \mathbf{B_y}(t) (\mathbf{u}_{ca} + \mathbf{\Delta u}) + \mathbf{D}.
\label{plant2_2}
\end{equation}
Writing $\mathbf{D}$ as $ \mathbf{D}=\mathbf{B_v} \mathbf{d}$, where the entries of the vector $\mathbf{d}$ is given in the Appendix A, expression in \eqref{plant2_2} can be rewritten as
\begin{equation}
\begin{aligned}
\dot{\mathbf{x}}&=\mathbf{A} \mathbf{x} + \mathbf{B_v} [\mathbf{B_{y}}(t) (\mathbf{u}_{ca} + \mathbf{\Delta u}) + \mathbf{d}] \\
&=\mathbf{A} \mathbf{x} + \mathbf{B_v} [\mathbf{B_{y}}(t) \mathbf{u}_{ca} + \mathbf{B_{y}}(t) \mathbf{\Delta u} + \mathbf{d}].
\end{aligned}
\label{plant2_3}
\end{equation}
Defining
\begin{equation}
\bar{\mathbf{d}} \equiv \mathbf{B_{y}}(t) \mathbf{\Delta u} + \mathbf{d},
\label{d_bar}
\end{equation}
\eqref{plant2_3} can be rewritten as
\begin{equation}
\dot{\mathbf{x}}=\mathbf{A} \mathbf{x} + \mathbf{B_v} [\mathbf{B_{y}}(t)\mathbf{u}_{ca} + \bar{\mathbf{d}}].
\label{disturbed_system}
\end{equation}

The time-varying matrix $\mathbf{B_{y}}(t)$ can be written as a multiplication of a constant matrix, $\mathbf{B_l}$, and a time-varying invertible matrix $\mathbf{B_n}(t)$, as $\mathbf{B_{y}}(t)=\mathbf{B_l} \mathbf{B_n}(t) $, where the contents of the matrices, $\mathbf{B_l} \in R^{5 \times 12}$ and $\mathbf{B_n}(t) \in R^{12 \times 12}$ are given in Appendix A. 
Then, this multiplication can be  substituted into \eqref{disturbed_system} to obtain
\begin{equation}
\dot{\mathbf{x}}=\mathbf{A} \mathbf{x} + \mathbf{B_v} [ \mathbf{B_l} \mathbf{B_n}(t) \mathbf{u}_{ca} + \bar{\mathbf{d}}].
\label{plant2_4}
\end{equation}
Defining $\mathbf{\bar{u}} \equiv \mathbf{B_n}(t) \mathbf{u}_{ca}$, and introducing a diagonal unknown matrix $\mathbf{\Lambda} \in R^{12 \times 12}$ with positive entries, the state equation given in \eqref{plant2_4} can be rewritten as
\begin{eqnarray}
\dot{\mathbf{x}}&=&\mathbf{A} \mathbf{x} +\mathbf{B_v} ( \mathbf{B_l}\mathbf{\Lambda} \mathbf{\bar{u}} + \bar{\mathbf{d}}) \\
\label{plant3_2}
&=&\mathbf{A} \mathbf{x} + \mathbf{B_v} \mathbf{v},
\label{plant3_3}
\end{eqnarray}
where $\mathbf{v} \in R^5$ is the virtual control input consisting of the desired forces and moments derived in Section \ref{sec:virtual-control-input-generation}. 
It is noted that faults and uncertainties in actuator effectiveness can be represented with the help of $\Lambda$. 
The objective of the control allocation is to determine the vector $\mathbf{u}$ to achieve
\begin{equation}
\mathbf{B_l} \mathbf{\Lambda} \mathbf{\bar{u}} + \bar{\mathbf{d}} =\mathbf{v},
\label{objective_2}
\end{equation}
where $\mathbf{B_l} \mathbf{\Lambda} \mathbf{\bar{u}} + \bar{\mathbf{d}}$ represents the net forces  moments acting on the vehicle, which can be measured using an inertial measurement unit. 

In order to build the mechanism for control allocation, we consider a dynamic system
\begin{equation}
\dot{\mathbf{\xi}}=\mathbf{A_m} \mathbf{\xi} + \mathbf{B_l} \mathbf{\Lambda} \mathbf{\bar{u}} + \bar{\mathbf{d}} - \mathbf{v},
\label{y1_2}
\end{equation}
where $\mathbf{A_m} \in R^{5 \times 5}$ is stable. A reference model can also be defined as
\begin{equation}
\dot{\mathbf{\xi}}_m=\mathbf{A_m} \mathbf{\xi_m}.
\label{reference_2}
\end{equation} 
$\mathbf{\bar{u}}$ can be created as $\mathbf{\bar{u}}=\boldsymbol{\theta}^T \mathbf{v}$, where $\boldsymbol{\theta} \in R^{12 \times 5}$ is a parameter matrix to be determined, and (\ref{y1_2}) can be rewritten as
\begin{equation}
\dot{\mathbf{\xi}}= \mathbf{A_m} \mathbf{\xi} + (\mathbf{B_l} \mathbf{\Lambda} \boldsymbol{\theta}^T - \mathbf{I})\mathbf{v} + \bar{\mathbf{d}}.
\label{y2_2}
\end{equation} 

Assuming an ideal parameter vector $\boldsymbol{\theta}^*$ that satisfies $\mathbf{B_l} \mathbf{\Lambda} \boldsymbol{\theta}^{*T} = \mathbf{I}$ and defining $\tilde{\boldsymbol{\theta}}^T \equiv \boldsymbol{\theta}^T - \boldsymbol{\theta}^{*T} $, where $\tilde{\boldsymbol{\theta}}$ is the deviation of the parameter vector $\boldsymbol{\theta}$ from its ideal value $\boldsymbol{\theta}{*}$, (\ref{y2_2}) can be rewritten as 
\begin{equation}
\dot{\mathbf{\xi}}=\mathbf{A_m} \mathbf{\xi} + \mathbf{B_l} \mathbf{\Lambda} \tilde{\boldsymbol{\theta}}^T \mathbf{v} + \bar{\mathbf{d}}.
\label{y3_2}
\end{equation} 
Defining an error as  $\mathbf{e}=\mathbf{\xi}-\mathbf{\xi_m}$ and subtracting (\ref{reference_2}) from (\ref{y3_2}), it is obtained that
\begin{equation}
\dot{\mathbf{e}}= \mathbf{A_m} \mathbf{e} + \mathbf{B_l} \mathbf{\Lambda} \tilde{\boldsymbol{\theta}}^T \mathbf{v} + \bar{\mathbf{d}}.
\label{error_2}
\end{equation} 
$\textbf{Theorem 1: }$ If the parameter matrix $\theta$ is updated with the adjustment law as,
\begin{equation}
\dot{\boldsymbol{\theta}}=\boldsymbol{\Gamma} Proj(\boldsymbol{\theta}, -\mathbf{v}\mathbf{e}^T \mathbf{P}\mathbf{B_l}),
\label{adaptivelaw_2}
\end{equation}
where $\boldsymbol{\Gamma} =\boldsymbol{\Gamma}^T=\gamma \mathbf{I_r} \in R^{5 \times 5} >0 $, $\gamma$ is a positive scalar, $\mathbf{I_r}$ is an identity matrix, $Proj$ refers to the projection operator \cite{Lavretsky2011}, and $\mathbf{P}$ is the positive definite symmetric solution of the Lyapunov equation $\mathbf{A_m}^T \mathbf{P}+ \mathbf{P}\mathbf{A_m}=-\mathbf{Q}$, where $\mathbf{Q}$ is a positive definite symmetric matrix, all the signals in the proposed control allocation structure given in \eqref{y1_2}-\eqref{error_2} and \eqref{objective_2} is achieved.  

$\textbf{Proof: }$See  \cite{Tohidi2016}, \cite{TOHIDI20175492} and  \cite{tohidinew}

$\textbf{Theorem 2: }$ The closed loop system, consisting of the plant \eqref{plant3_3}, and the control input, $\mathbf{v}$, whose elements are given in \eqref{fc2} - \eqref{Fy_law}, is stable.

$\textbf{Proof: }$ See Appendix B.
\section{SIMULATIONS}
\begin{figure*}
    \begin{center}
        \begin{subfigure}{\linewidth}
            \includegraphics[trim={5cm 4cm 6cm 3.5cm},clip=true,width=\linewidth, height=7.7cm, page=8]{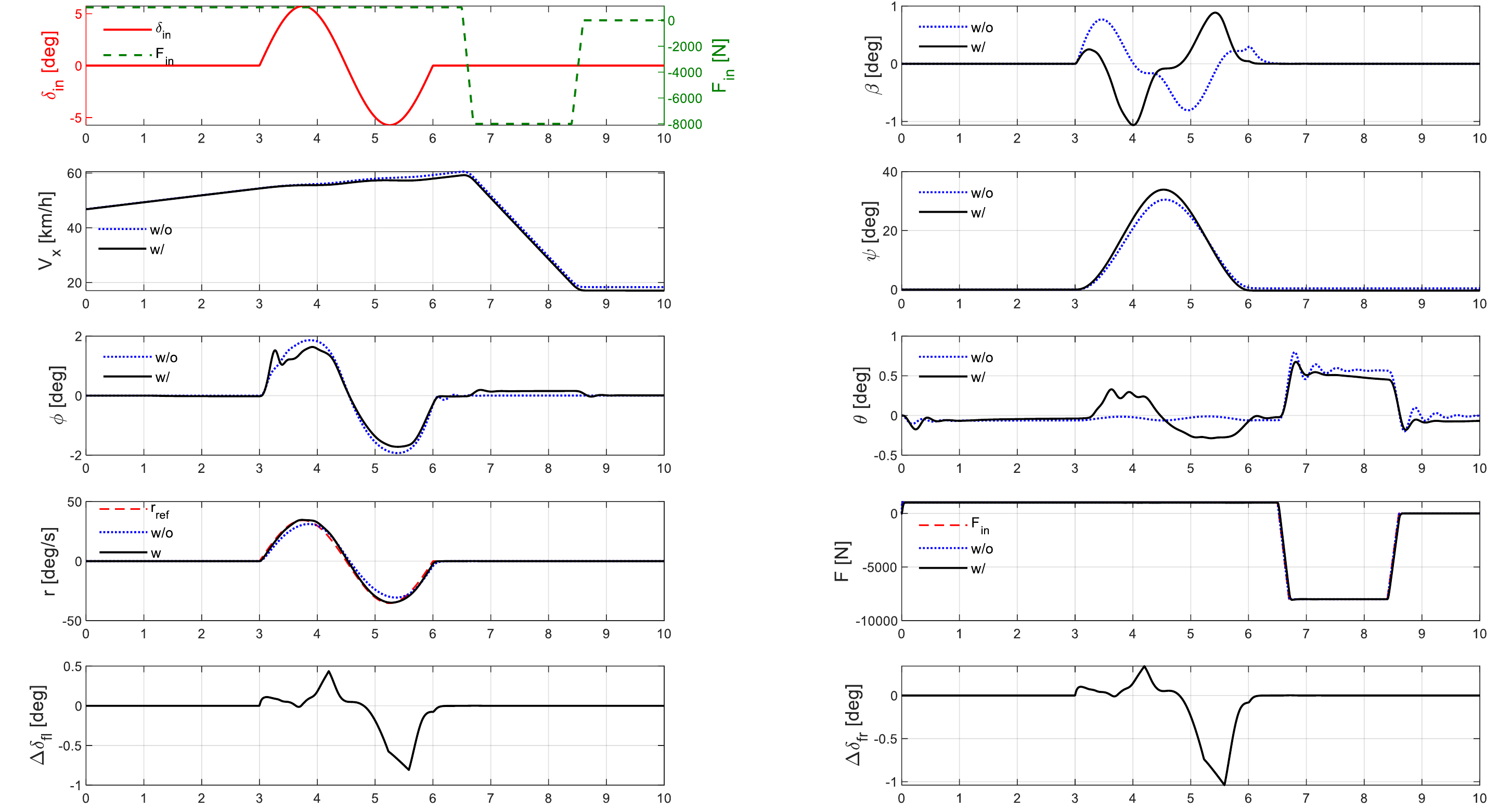}
        \end{subfigure}    
        
        \begin{subfigure}{\linewidth}
            \includegraphics[trim={5cm 4cm 5cm 3.5cm},clip=true,width=\linewidth, height=7.7cm, page=9]{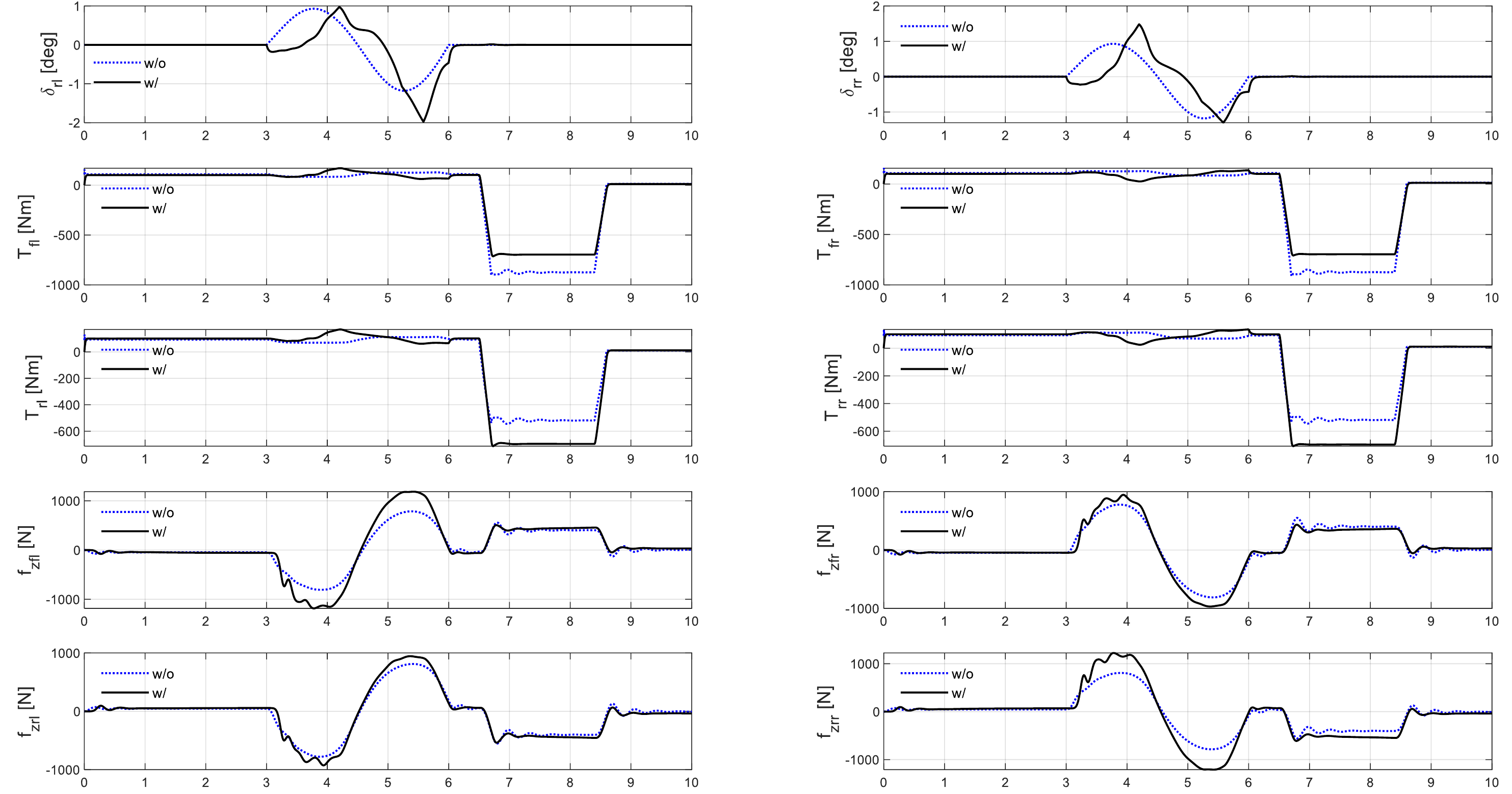}
        \end{subfigure}    
        
        \begin{subfigure}{\linewidth}
            \includegraphics[trim={5cm 14cm 5cm 13.5cm},clip=true,width=\linewidth, height=2.5cm, page=10]{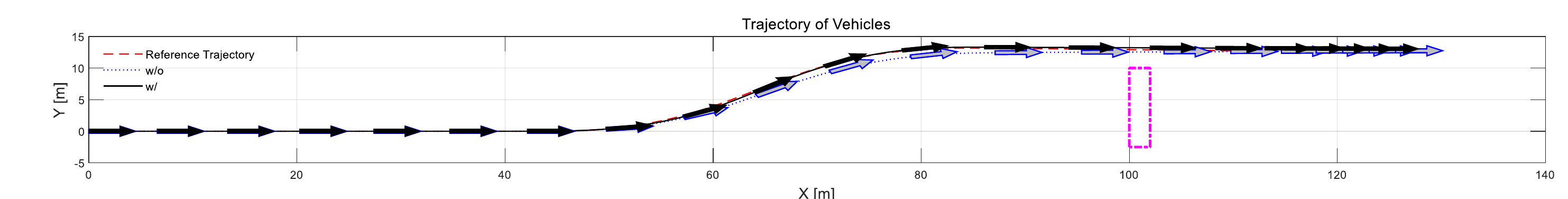}
        \end{subfigure}
    \end{center}
    \caption{Object Avoidance Maneuver in Low Velocity Scenario. "w/" and "w/o" refer to the cases with and without the proposed control framework, respectively.}
    \label{low} 
\end{figure*}
\begin{figure*}
    \begin{center}
        \begin{subfigure}{\linewidth}
            \includegraphics[trim={5cm 4cm 6cm 3.5cm},clip=true,width=\linewidth, height=7.7cm, page=12]{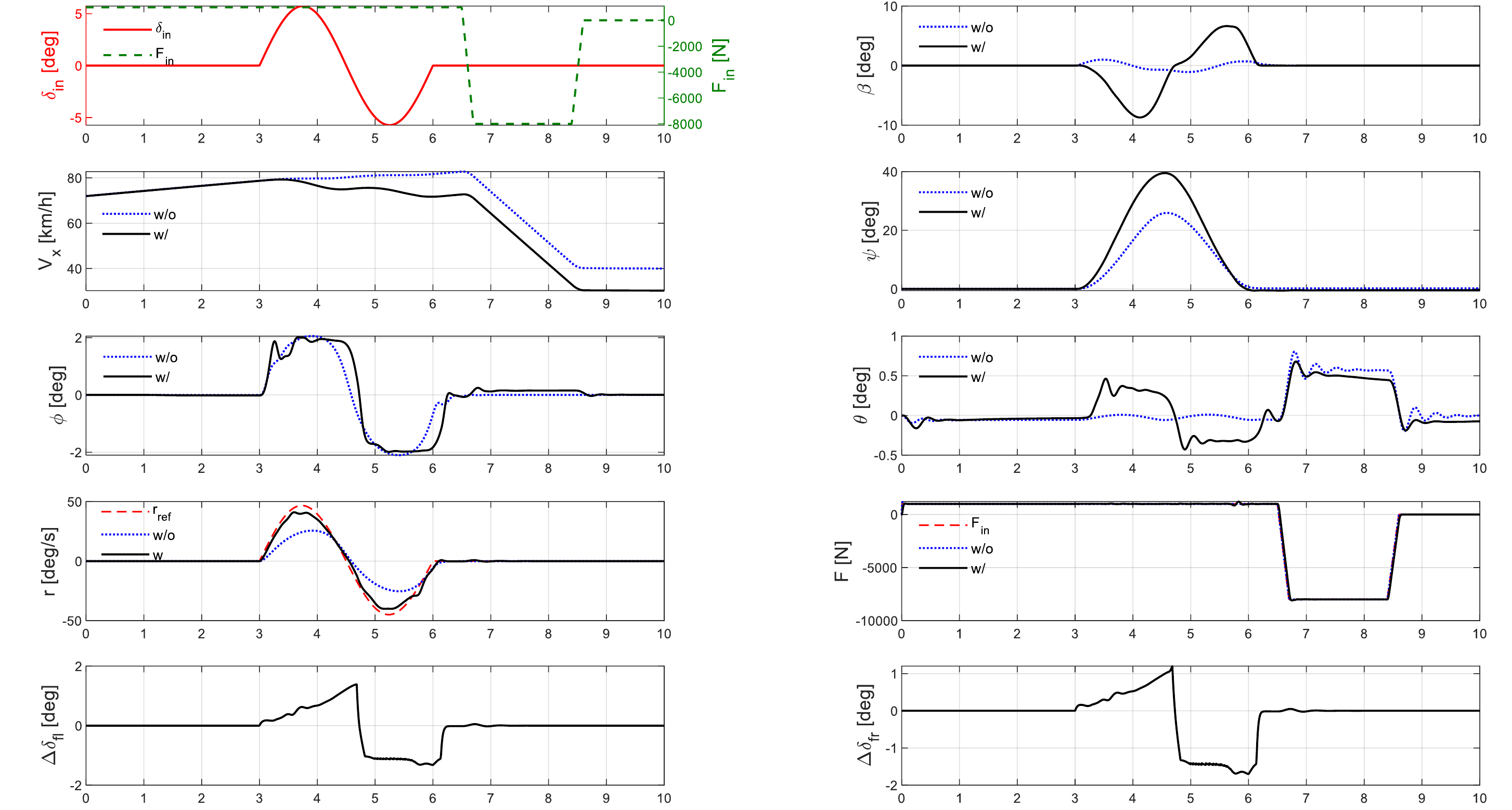}
        \end{subfigure}    
        
        \begin{subfigure}{\linewidth}
            \includegraphics[trim={5cm 4cm 5cm 3.5cm},clip=true,width=\linewidth, height=7.7cm, page=13]{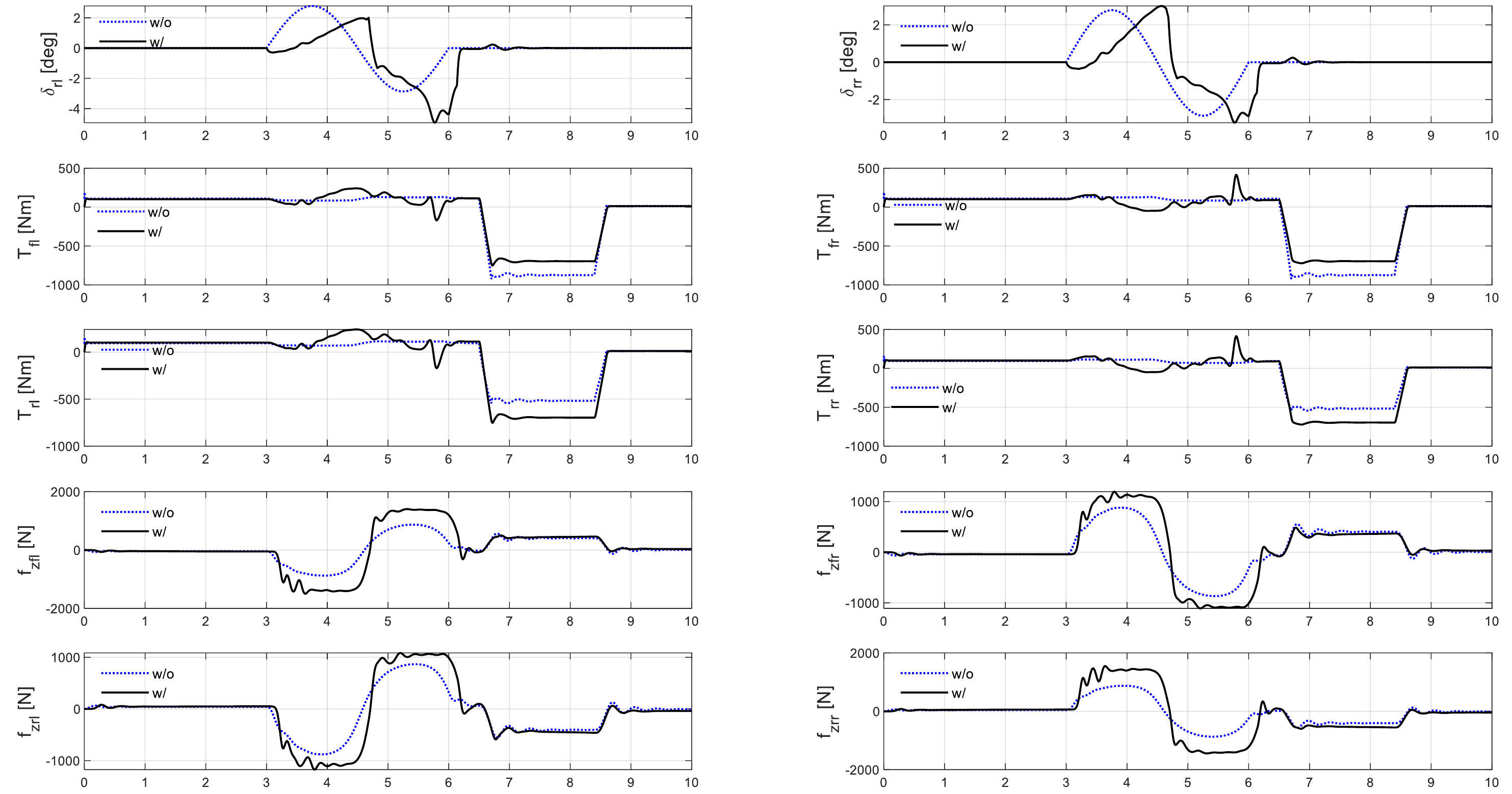}
        \end{subfigure}    
        
        \begin{subfigure}{\linewidth}
            \includegraphics[trim={5cm 13.5cm 5cm 13.5cm},clip=true,width=\linewidth, height=2.5cm, page=14]{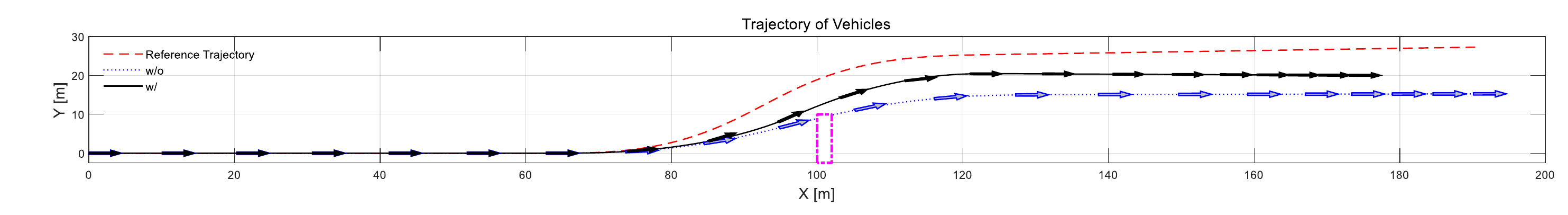}
        \end{subfigure}
    \end{center}
    \caption{Object Avoidance Maneuver in High-Velocity Scenario. "w/" and "w/o" refer to the cases with and without the proposed control framework, respectively.}
    \label{high} 
\end{figure*}
\begin{figure*}
    \begin{center}
        \begin{subfigure}{\linewidth}
            \includegraphics[trim={5cm 4cm 6cm 3.5cm},clip=true,width=\linewidth, height=7.7cm, page=16]{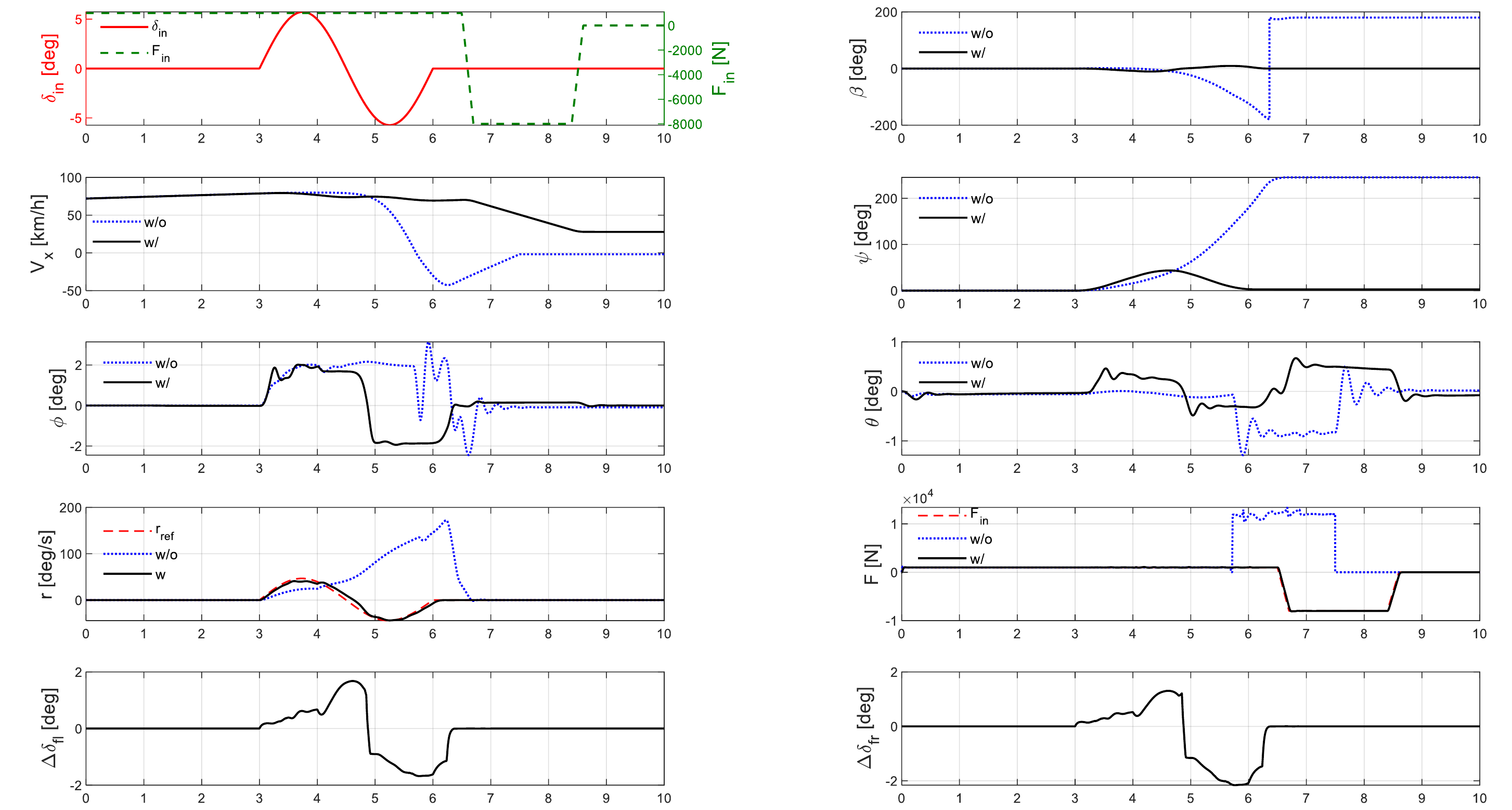}
        \end{subfigure}    
        
        \begin{subfigure}{\linewidth}
            \includegraphics[trim={5cm 4cm 5cm 3.5cm},clip=true,width=\linewidth, height=7.7cm, page=17]{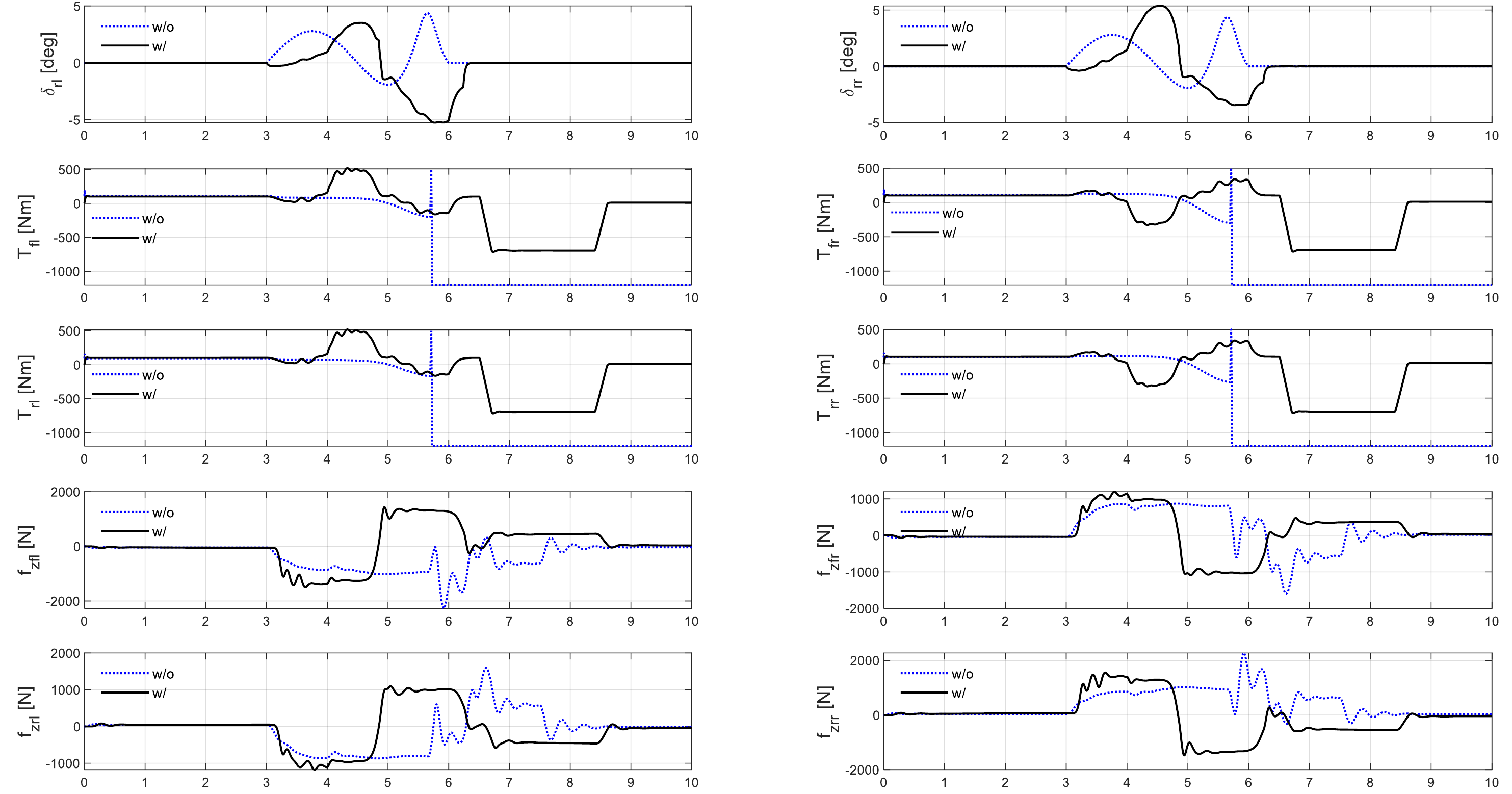}
        \end{subfigure}    
        
        \begin{subfigure}{\linewidth}
            \includegraphics[trim={5cm 13.5cm 5cm 13.5cm},clip=true,width=\linewidth, height=2.5cm, page=18]{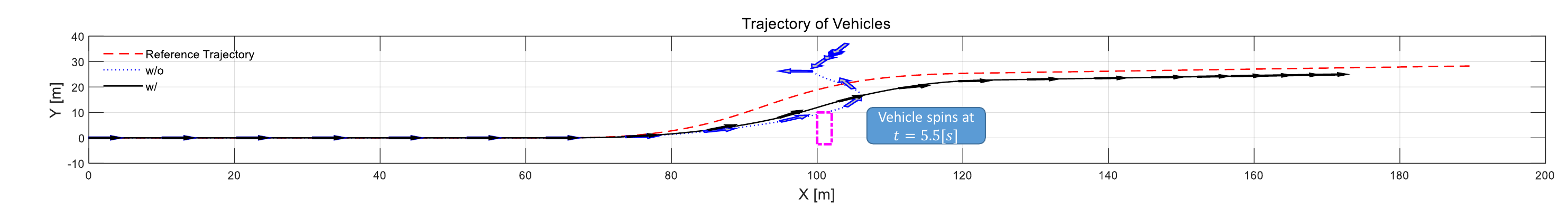}
        \end{subfigure}
    \end{center}
    \caption{Object Avoidance Maneuver with Varying Road Conditions. "w/" and "w/o" refer to the cases with and without the proposed control framework, respectively.}
    \label{vary} 
\end{figure*}
\begin{figure*}
    \begin{center}
        \begin{subfigure}{\linewidth}
            \includegraphics[trim={5cm 4cm 6cm 3.5cm},clip=true,width=\linewidth, height=7.6cm, page=20]{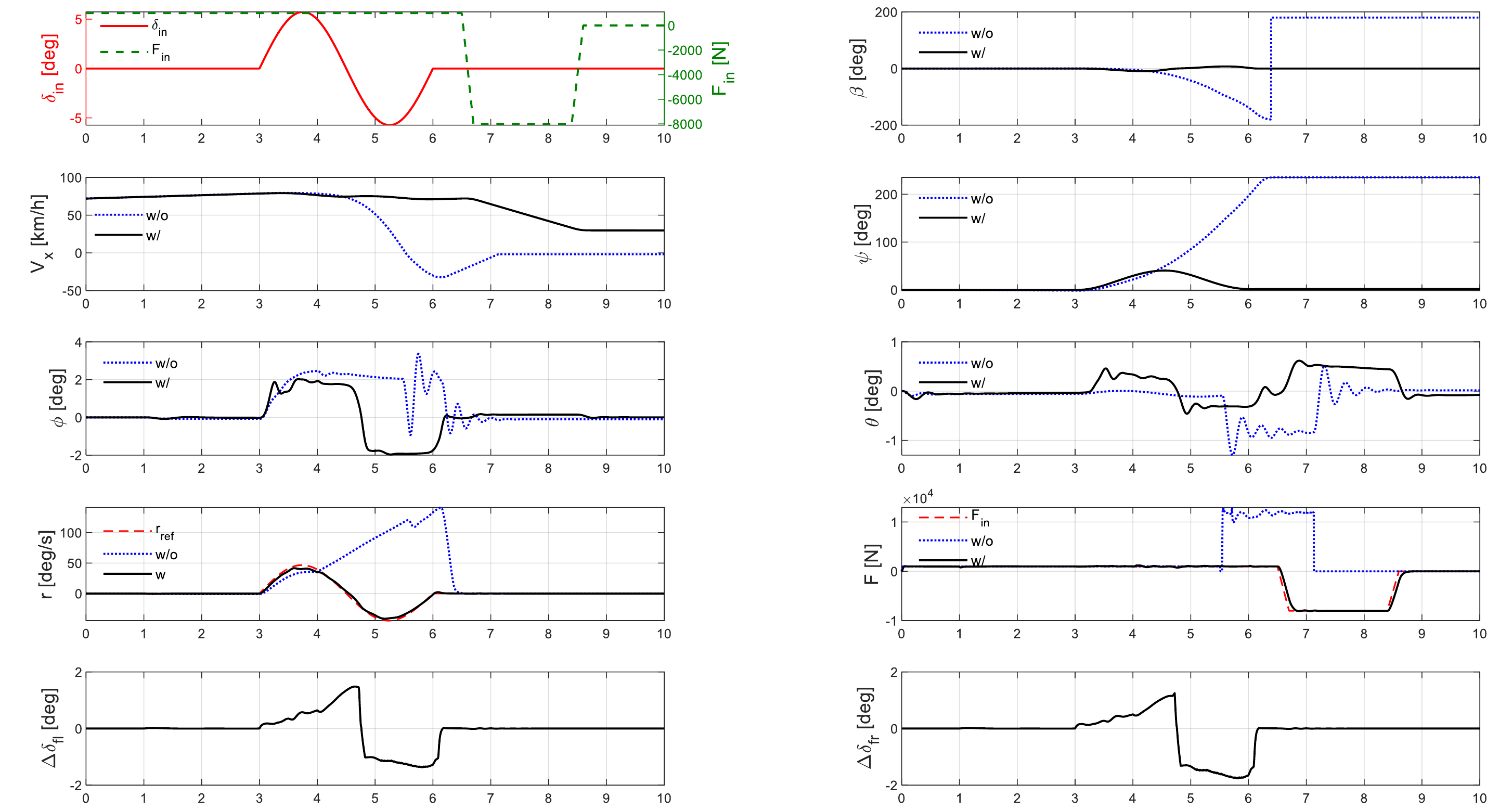}
        \end{subfigure}    
        
        \begin{subfigure}{\linewidth}
            \includegraphics[trim={5cm 4cm 5cm 3.5cm},clip=true,width=\linewidth, height=7.6cm, page=21]{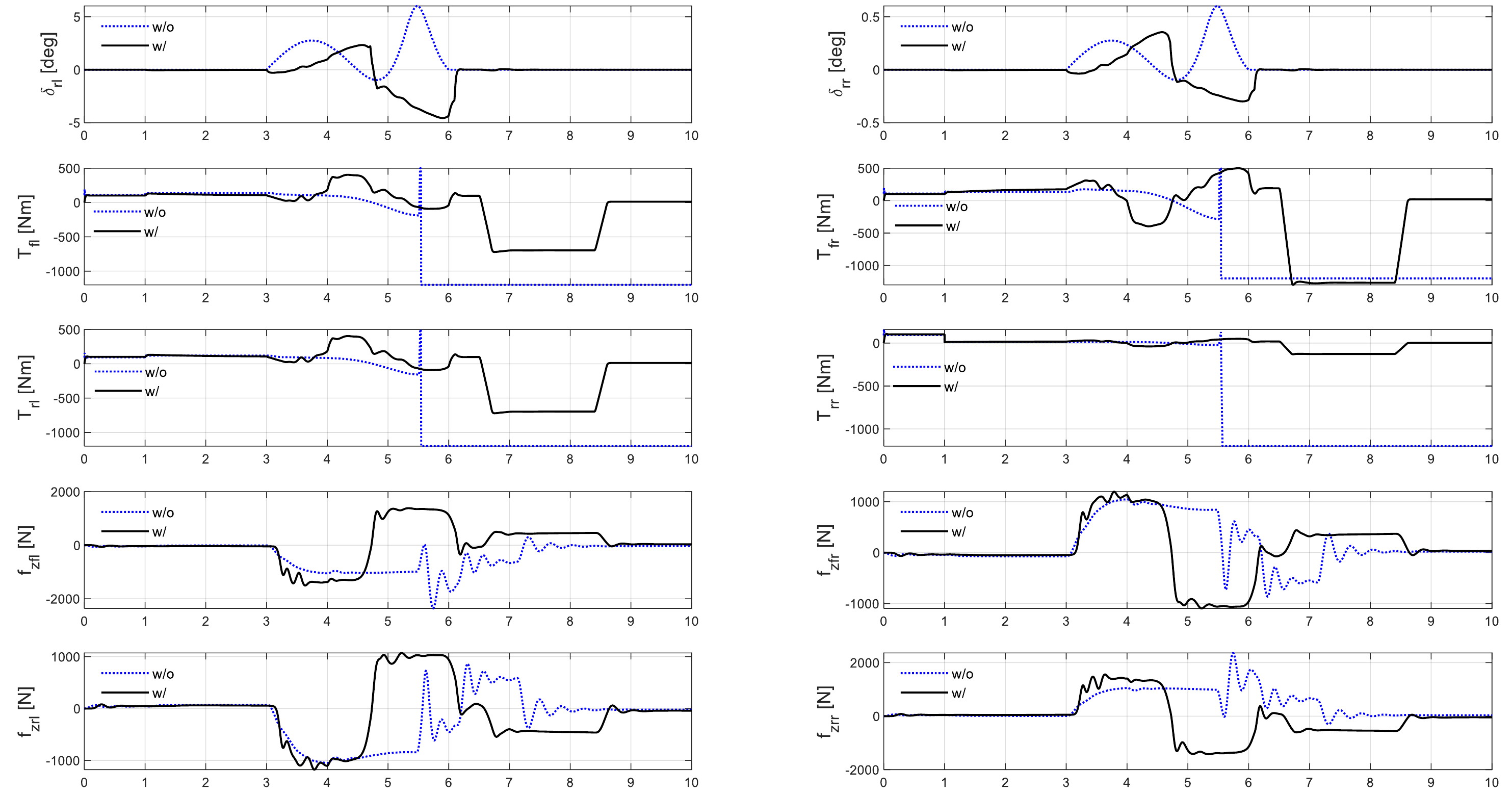}
        \end{subfigure}    
        
        \begin{subfigure}{\linewidth}
            \includegraphics[trim={5cm 13.5cm 5cm 13.5cm},clip=true,width=\linewidth, height=2.4cm, page=22]{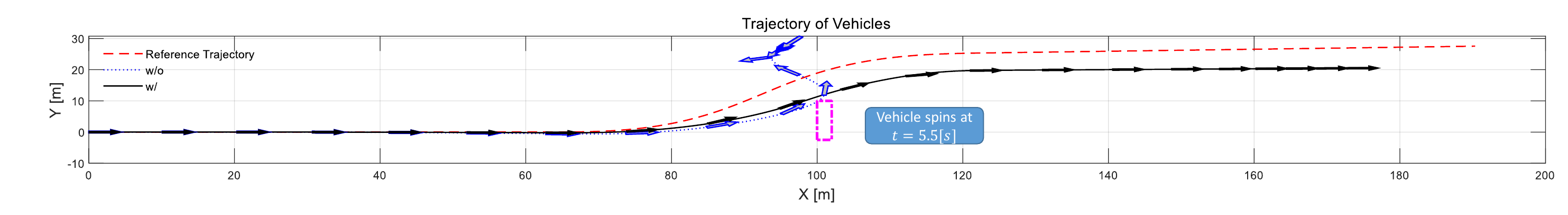}
        \end{subfigure}
    \end{center}
    \caption{Object Avoidance Maneuver with Actuator Failure and Varying Road Conditions. "w/" and "w/o" refer to the cases with and without the proposed control framework, respectively.}
    \label{fail} 
\end{figure*}
\begin{figure*}
    \begin{center}
        \includegraphics[trim={5cm 4cm 5cm 4.5cm},clip=true,width=\linewidth, height=7.9cm, page=26]{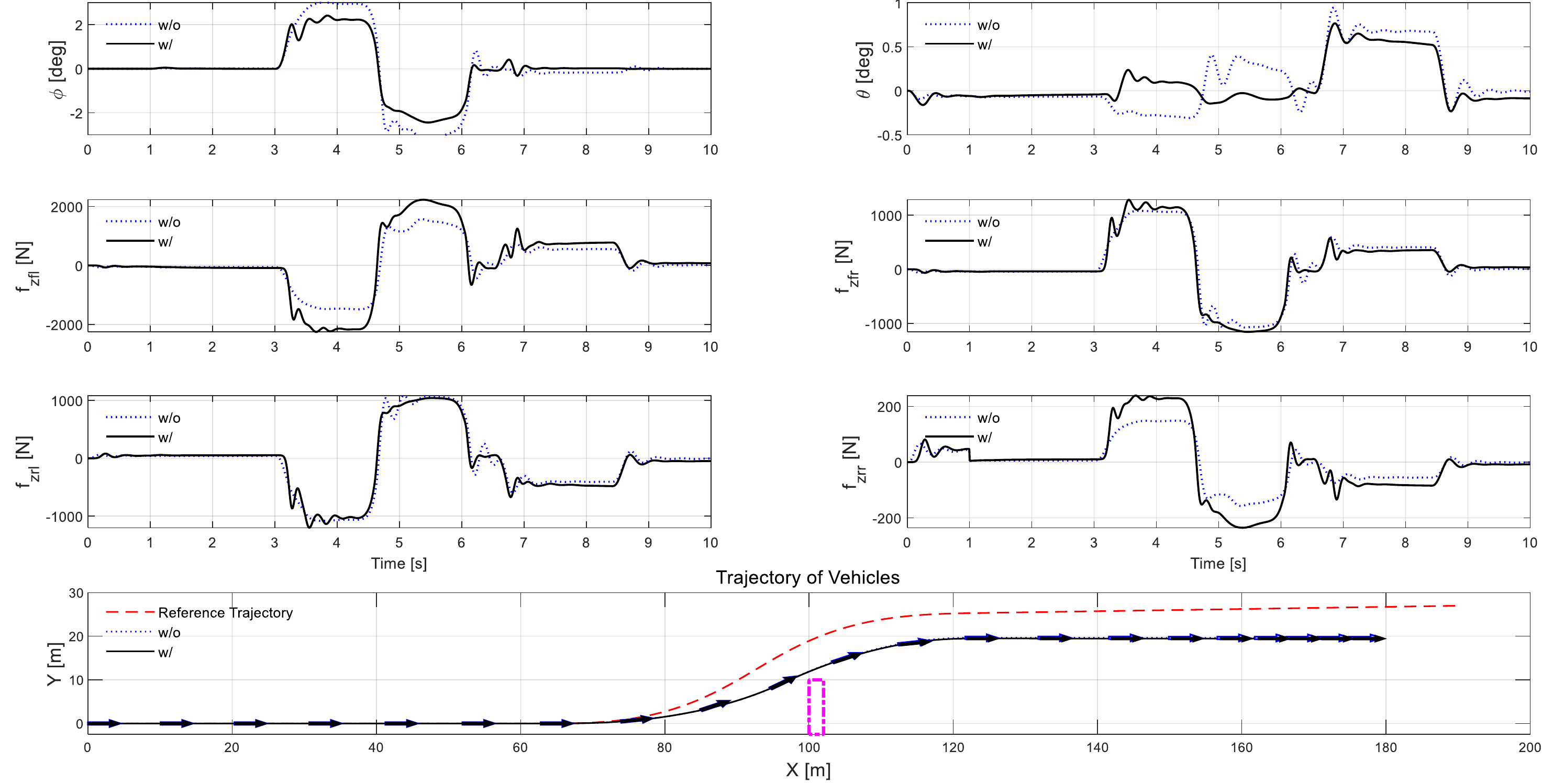}
    \end{center}
    \caption{Comparison of Active Suspensions in Case of Failure. "w/" and "w/o" refer to the cases with and without the proposed control framework, respectively.}
    \label{suspension} 
\end{figure*}
In order to validate the proposed control framework, a Matlab/Simulink model of the overall closed-loop system is constructed. Different failure scenarios are simulated, and the results are compared with a baseline controller.
\subsection{Baseline Controller}
The baseline controller has three separate subsystems consisting of rear-wheel steering control, traction force control, and active suspension control.
\subsubsection{Rear Wheel Steering Control}
In the baseline controller, the rear wheel's angle of rotation is determined to be proportional to the front wheels' rotation angle. This proportion is determined based on the studies presented in \cite{Marino2007, Ulsoy2012}, where the amount of the gain, $K_s$, between the front and rear-wheel steering is tuned to improve stability and performance: At low speeds, $K_s$ is negative to increase maneuverability. At high speeds, however, the gain is positive, which increases stability. Specifically, $K_s$ is determined as
\begin{equation}
K_s= \frac{\delta_r}{\delta_f}=\frac{m V_x^2 a-b L C_{\alpha} N_r}{m V_x^2 b+a L C_{\alpha} N_f} \cdot \frac{C_{\alpha} N_f}{C_{\alpha} N_r},
\label{base1}
\end{equation}
where $L$ is the length of the wheelbase, $m$ is the vehicle mass, $V_x$ is the longitudinal velocity, $C_{\alpha}$ is the lateral friction coefficient, $N_f$, $N_r$ are normal forces at the front and rear tires, respectively and $\delta_f$ and $\delta_r$ are rear and front steering angles.

\subsubsection{Traction Control}
The total traction force, $F_c$, is generated using a PI controller similar to the one used in the proposed virtual control input given in \eqref{fc2}. Then, this total traction force is distributed among the wheels proportional to the normal forces at the tires. 
The resulting control law can be written as
\begin{equation}
T_{ij}=\frac{F_c \ mg}{4N_{ij}},
\label{base2}
\end{equation}
where $mg$ is the weight of the vehicle, and $T_{ij}$ and $N_{ij}$ are the torque and the normal force at wheel $ij$, $i \in \{f,r\} $, $j \in \{l,\bar{r}\}$.

\subsubsection{Active Suspension Control}
For active suspension control, a similar approach given in \cite{876977} is used. 
Active suspension forces are determined by two PI controllers to stabilize the roll and pitch motions. 
PI controllers use the deviations of the roll, $\theta$, and pitch, $\phi$, angles to determine the required active suspension forces for stabilization. 
These forces are then distributed to the individual wheels, according to their moment creation effects provided in \eqref{pitch1} and \eqref{roll}, as
\begin{equation}
\begin{aligned}
f_{pitch} &= -K_{pp} \theta - K_{ip} \int \theta dt \\
f_{roll} &=  -K_{pr} \phi - K_{ir} \int \phi dt
\end{aligned}
\label{base3}
\end{equation}
\begin{equation}
\begin{aligned}
f_{zfl} &= - f_{pitch} + f_{roll}  \\
f_{zf\bar{r}} &= - f_{pitch} - f_{roll}  \\
f_{zrl} &=  f_{pitch} + f_{roll}  \\
f_{zr\bar{r}} &=  f_{pitch} - f_{roll},
\end{aligned}
\label{base4}
\end{equation}
where $f_{zij}$, $i \in \{f,r\} $, $j \in \{l,\bar{r}\}$ is the active suspension force at wheel $ij$.
\subsection{ Low-Speed Performance}
Initially, the proposed control framework is compared to the baseline controller for the case of no failure and when the initial velocity of the vehicle is set as $V_x=13$ m/s. The scenario consists of an object avoidance maneuver followed by an emergency braking. 
The simulation results are given in Fig. \ref{low}, where the bottom sub-figure shows the trajectories of the vehicles with the baseline (w/o) and the proposed controllers (w/). In the sub-figure, the obstacle is represented by a pink dash-dotted line at $x=100$ m. 
In this scenario, the steering maneuver starts at $t=3$ s and ends at $t=6$ s. Later, at $t=6.5$ s driver brakes for $1$ s and then driver gives no throttle or brake input. 
Fig. \ref{low} also shows the allocated signals, which are steering angle corrections, $\Delta \delta_{ij} $, wheel torques $T_{ij}$, active suspension forces, $f_{zij}$, $i \in \{f,r\} $, $j \in \{l,\bar{r}\}$, and vehicle states $\mathbf{x} = \begin{bmatrix} \beta & V_x & \psi & \phi & \theta \end{bmatrix}$. 
As seen from the figure, the proposed system performs similarly to the baseline system under low speed and no-fault conditions.
\subsection{High-Speed Performance}
In this scenario, the proposed controller is compared to the baseline system when the initial velocity, $V_x$, is $20$ m/s. 
The same steering and acceleration/brake inputs are applied as the previous case, and there are no failures in the system. The simulation results are given in Fig. \ref{high}.
When side-slip, $\beta$, values of the two vehicles are compared, it is seen that the baseline vehicle has much less side-slip. 
This is due to the nature of the rear steering gain $K_s$, which is designed to maintain yaw stability. 
However, this reduces yaw rotation and prevents the vehicle from escaping the obstacle. On the other hand, the proposed control framework provides a better yaw rate reference following while also keeping the vehicle stable.
\subsection{Performance in Varying Road Conditions}
In this simulation, the road friction coefficient $C_{\alpha}$ for the right tires is reduced to $60\%$ of its original value to simulate slippery road conditions, starting at $t=4$ s. The vehicle is commanded to perform the same object avoidance maneuver with an initial velocity of 20 m/s.
The results in Fig. \ref{vary} show that the baseline vehicle has difficulty following the commands while the proposed controller performs as designed. 
In order to compensate for varying road conditions, the proposed system shifts torque distribution toward the right wheels.
Moreover, trajectories, side slip angles, and velocities show that the vehicle equipped with the proposed controller remains stable, while the baseline system over-steers at the first steering input and spins out of control.
\subsection{Actuator Failure with Varying Road Conditions}
In order to test the performance of the proposed controller in challenging situations, an ``effectiveness loss at the wheel'' scenario is created. 
In this scenario, the rear right tire traction force and steering angle are reduced to $10\%$ of their original values, at $t=1$ s. 
Additionally, at $t=4$ s, the lateral friction coefficient is reduced by $10\%$. In the scenario, the initial velocity is also set to 20 m/s.
The results of the simulation are given in Fig. \ref{fail}. 
The proposed controller modifies the steering angles to compensate for the undesired moments created by actuator effectiveness losses, which ensures a stable turn around the obstacle and a proper following of the reference trajectory. On the other hand,  the baseline system cannot make a stable turn and spins. (It is observed that even for a lower initial velocity of 15 m/s, the baseline system loses stability.)
\subsection{Integrated vs Independent Active Suspension}
To demonstrate the advantages of an integrated active suspension system, the suspension control in the proposed control framework is replaced with that of the baseline controller, and the resulting structure is compared with the original integrated framework. The outcomes are presented in Fig. \ref{suspension}.
In order to level the comparison, controller parameters are adjusted such that when there is no fault, the system performances are similar.
In this scenario, the initial velocity is set to 20 m/s, and the effectiveness of the suspension actuators at the rear right wheel is reduced to $10 \%$ of its original value when $t = 1$ s.
The proposed control framework increases the suspension force in the front left wheel $f_{zl}$, to compensate for the failure. 
As a result, despite the initial adaptation phase, the proposed scheme keeps the roll and pitch angles approximately $35\%$ lower, compared to the case where the baseline controller is used for suspension control.
It is noted that smaller roll and pitch angles result in a reduced shift in the center of gravity and a more stable and comfortable ride is achieved \cite{Ulsoy2012}.
\section{CONCLUSIONS}
In this paper, a novel control framework is introduced as an integrated vehicle controller that can handle the uncertainties and non-linearities of the lateral and vertical dynamics of the vehicle motion. 
The framework is designed based on and validated by a 14 degrees of freedom model that incorporates steering, suspension, and forward motion dynamics.
The controller employs steering, traction, and suspension forces to follow desired yaw and force references while keeping the vehicle stable under unfavorable driving conditions. 
Simulation results show that the presented framework provides a comparable performance with the baseline system in a low-velocity and no-failure scenario. 
However, when the velocity is higher, the baseline system cannot perform as well as the proposed control framework. 
Moreover, when a fault or a road uncertainty is introduced, the proposed system can tolerate these non-ideal situations (up to $30\%$ higher longitudinal maneuver velocity), whereas the baseline controller oversteers and makes the vehicle spin.
Finally, our work shows that an integrated active suspension is better at overcoming a difficult maneuver with a better ride performance (approximately $35\%$ lower roll and pitch angles during same steering conditions) compared to an independent suspension control.


\medskip
\bibliographystyle{IEEEtran}
\bibliography{Journal}
\section{APPENDIX}
\subsection{Matrices for Control Allocation}
\begin{equation}
\mathbf{d}^T=
\begin{bmatrix}
\frac{V_0^2 \rho C_d A}{2} & 0 & 0 & -\frac{V_0^2 \rho C_d A}{2} & 0
\end{bmatrix},
 \label{d}
\end{equation}
where $V_0$ is linearized longitudinal velocity constant.
\begin{equation}
\begin{aligned}
\mathbf{B_y}(t) &= 
\begin{split}
&[\begin{matrix}
\mathbf{B_{y1}} & \mathbf{B_{y2}} & \mathbf{B_{y3}} &\mathbf{B_{y4}} & \mathbf{B_{y5}} & \mathbf{B_{y6}} \end{matrix} \\
&\begin{matrix} \mathbf{B_{y7}} & \mathbf{B_{y8}} & \mathbf{B_{y9}} & \mathbf{B_{y10}} & \mathbf{B_{y11}} & \mathbf{B_{y12}} \end{matrix}].
\end{split}
\end{aligned}
\label{B(t)}
\end{equation}
where
\footnotesize
\begin{equation}
\begin{aligned}
\mathbf{B_{y1}}&= \begin{bmatrix} 0 & C_{\alpha} N_{fl}(t) \cos(\delta_{fl}) & C_{\alpha} a N_{fl}(t) \cos(\delta_{fl}) & 0 & 0 \end{bmatrix}^T\\
\mathbf{B_{y2}}&= \begin{bmatrix} 0 & C_{\alpha} N_{fr}(t) \cos(\delta_{f\bar{r}}) & C_{\alpha} a N_{fr}(t) \cos(\delta_{f\bar{r}}) & 0 & 0 \end{bmatrix}^T\\
\mathbf{B_{y3}}&= \begin{bmatrix} 0 & C_{\alpha} N_{rl}(t) \cos(\delta_{rl}) & -C_{\alpha} b N_{rl}(t) \cos(\delta_{rl}) & 0 & 0 \end{bmatrix}^T\\
\mathbf{B_{y4}}&= \begin{bmatrix} 0 & C_{\alpha} N_{rr}(t) \cos(\delta_{r\bar{r}}) & -C_{\alpha} b N_{rr}(t) \cos(\delta_{r\bar{r}}) & 0 & 0 \end{bmatrix}^T\\
\mathbf{B_{y5}}&= \begin{bmatrix} \frac{\cos(\delta_{fl})}{R} & 0 & \frac{w \cos(\delta_{fl})}{2} & 0 & 0 \end{bmatrix}^T \\
\mathbf{B_{y6}}&= \begin{bmatrix} \frac{\cos(\delta_{f\bar{r}})}{R} & 0 & -\frac{w \cos(\delta_{f\bar{r}})}{2} & 0 & 0 \end{bmatrix}^T \\
\mathbf{B_{y7}}&= \begin{bmatrix} \frac{\cos(\delta_{rl})}{R} & 0 & \frac{w \cos(\delta_{rl})}{2} & 0 & 0 \end{bmatrix}^T \\
\mathbf{B_{y8}}&= \begin{bmatrix} \frac{\cos(\delta_{r\bar{r}})}{R} & 0 & -\frac{w \cos(\delta_{r\bar{r}})}{2} & 0 & 0 \end{bmatrix}^T \\
\mathbf{B_{y9}}&= \begin{bmatrix} 0 & 0 & 0 & w/2 & -a  \end{bmatrix}^T \\
\mathbf{B_{y10}}&= \begin{bmatrix} 0 & 0 & 0 & w/2 & -a \end{bmatrix}^T \\
\mathbf{B_{y11}}&= \begin{bmatrix} 0 & 0 & 0 & -w/2 & b \end{bmatrix}^T \\
\mathbf{B_{y12}}&= \begin{bmatrix} 0 & 0 & 0 & w/2 & b \end{bmatrix}^T,
\end{aligned}
\label{B(t)1}
\end{equation}
\normalsize
where $C_{\alpha}$ is the linearized friction coefficient. 
\begin{equation}
\begin{aligned}
\mathbf{B_n} =
\begin{split}
diag \bigg\{&
\frac{4N_{fl}}{m} cos(\delta_{fl}), \  \frac{N_{fr}}{m} cos(\delta_{f\bar{r}}),  \  \\ 
& \frac{4N_{rl}}{m} cos(\delta_{rl}), \  \frac{4N_{rr}}{m} cos(\delta_{r\bar{r}}), \  cos(\delta_{fl}),  \\
& cos(\delta_{f\bar{r}}), \   cos(\delta_{rl}), \  cos(\delta_{r\bar{r}}), \  1, \  1, \  1, \  1 \bigg\}.
\end{split}
\end{aligned}
\label{B_n}
\end{equation}
\begin{equation}
\begin{aligned}
\mathbf{B_l} &= 
\begin{split}
&[\begin{matrix}
\mathbf{B_{l_1}} & \mathbf{B_{l_2}} & \mathbf{B_{l_3}} & \mathbf{B_{l_4}} & \mathbf{B_{l_5}} & \mathbf{B_{l_6}} \end{matrix} \\
&\begin{matrix} \mathbf{B_{l_7}} & \mathbf{B_{l_8}} & \mathbf{B_{l_9}} & \mathbf{B_{l_{10}}} & \mathbf{B_{l_{11}}} &\mathbf{B_{l_{12}}} \end{matrix}],
\end{split}
\end{aligned}
\label{B_l}
\end{equation}
where 
\begin{equation}
\begin{aligned}
\mathbf{B_{l_1}}&= \begin{bmatrix} 0 & C_{\alpha} m/4 & a C_{\alpha} m/4 & 0 & 0 \end{bmatrix}^T \\
\mathbf{B_{l_2}}&= \begin{bmatrix} 0 & C_{\alpha} m/4 & a C_{\alpha} m/4 & 0 & 0\end{bmatrix}^T \\
\mathbf{B_{l_3}}&= \begin{bmatrix} 0 & C_{\alpha} m/4 & -b C_{\alpha} m/4 & 0 & 0\end{bmatrix}^T \\
\mathbf{B_{l_4}}&= \begin{bmatrix} 0 & C_{\alpha} m/4 & -b C_{\alpha} m/4 & 0 & 0\end{bmatrix}^T \\
\mathbf{B_{l_5}}&= \begin{bmatrix} 1/R & 0 & -w/2 & 0 & 0 \end{bmatrix}^T \\
\mathbf{B_{l_6}}&= \begin{bmatrix} 1/R & 0 & w/2 & 0 & 0 \end{bmatrix}^T \\
\mathbf{B_{l_7}}&= \begin{bmatrix} 1/R & 0 & -w/2 & 0 & 0 \end{bmatrix}^T \\
\mathbf{B_{l_8}}&= \begin{bmatrix} 1/R & 0 & w/2 & 0 & 0 \end{bmatrix}^T \\
\mathbf{B_{l_9}}&=    \begin{bmatrix} 0 & 0 & 0 & w/2 & -a \end{bmatrix}^T \\
\mathbf{B_{l_{10}}}&= \begin{bmatrix} 0 & 0 & 0 & -w/2 & -a \end{bmatrix}^T \\
\mathbf{B_{l_{11}}}&= \begin{bmatrix} 0 & 0 & 0 & w/2 & b \end{bmatrix}^T \\
\mathbf{B_{l_{12}}}&= \begin{bmatrix} 0 & 0 & 0 & -w/2 & b \end{bmatrix}^T.
\end{aligned}
\label{B_l1}
\end{equation}

\subsection{Close Loop Stability}
Consider the plant dynamics given in \eqref{plant3_3}. Using \eqref{fc2}, the Laplace transform of first element of the virtual control input $\mathbf{V}(s)$ can be written as
\begin{equation}
\mathbf{V}_1(s) = 
(F_{ref}(s) - F(s))(K_{pf}  + K_{if}/s),
\label{v1}
\end{equation}
where, $F(s)$ can be expressed as $F(s) = m a_x(s)$. Similarly, the second element of $\mathbf{V}(s)$ can be written, by using \eqref{Fy_law}, as
\begin{equation}
\mathbf{V}_2(s) = (\beta_{ref}(s) - \beta(s))(K_{py} + K_{iy} / s),
\end{equation}
where, side-slip angle $\beta$ can be expressed as $\beta(s) = V_y(s) / V_0$. Using \eqref{Mz}, the third element of $\mathbf{V}(s)$ can be obtained as
\begin{equation}
\begin{split}
\mathbf{V}_3(s) &= (r_{ref}(s) - r(s))(K_{pm} + K_{im} / s) \\
				&+ \beta(s) (K_{ps} + K_{is} / s).
\end{split}
\end{equation}
Finally, last two elements of $\mathbf{V}(s)$ can be obtained by taking the Laplace transform of \eqref{roll_pitch} as
\begin{equation}
\begin{gathered}
\mathbf{V}_4(s)= (\theta_{ref}(s) - \theta(s))(K_{pp} + K_{ip} / s + s K_{dp}) \\
\mathbf{V}_5(s)= (\phi_{ref}(s) - \phi(s)) (K_{pr} + K_{ir} / s + s K_{dr}).
\end{gathered}
\label{v5}
\end{equation}
Using \eqref{v1} - \eqref{v5}, $\mathbf{V}(s)$ can be expressed in vector form as
\footnotesize
\begin{equation}
\mathbf{V}(s) = 
\begin{bmatrix}
(F_{ref}(s) - m a_x (s))(K_{pf}  + K_{if}/s) \\
(\beta_{ref}(s) - \frac{V_y(s)}{V_0})(K_{py} + K_{iy} / s) \\
(r_{ref}(s) - r(s))(K_{pm} + K_{im} / s) + \frac{Vy(s)}{V_0} (K_{ps} + K_{is} / s)\\
(\theta_{ref}(s) - \theta(s))(K_{pp} + K_{ip} / s + s K_{dp}) \\
(\phi_{ref}(s) - \phi(s)) (K_{pr} + K_{ir} / s + s K_{dr})
\end{bmatrix},
\label{virtual_control_laplace}
\end{equation}
\normalsize
which can be rewritten as
\begin{equation}
\mathbf{V}(s) = \mathbf{K}(s) (\mathbf{R}(s) - \mathbf{Y}(s)),
\label{closeloop_stability0}
\end{equation}
where $\mathbf{R}(s)$ is the Laplace transform of the reference vector,
\begin{equation}
\mathbf{r}=
\begin{bmatrix}
 F_{ref} & \beta_{ref} & r_{ref} & \theta_{ref} & \phi_{ref}
 \end{bmatrix},
\label{ref_vec}
\end{equation}
$\mathbf{Y}(s)$ is the Laplace transform of the output vector
\begin{equation}
\mathbf{y}=
\begin{bmatrix}
 F & \beta & r & \theta & \phi
 \end{bmatrix},
\label{output_vec}
\end{equation}
and $\mathbf{K}(s)$ is the Laplace transform of the control matrix, whose rows are given as
\footnotesize
\begin{equation}
\begin{aligned}
\mathbf{K}_1(s) =& \begin{bmatrix} (K_{pf} + K_{if} / s) & 0 & 0 & 0 & 0  \end{bmatrix} \\
\mathbf{K}_2(s) =& \begin{bmatrix} 0 & (K_{py} + K_{iy} / s) & 0 & 0 & 0 \end{bmatrix} \\
\mathbf{K}_3(s) =& \begin{bmatrix} 0 & (K_{pb} + K_{ib} / s) & (K_{pmz} + K_{imz} / s) & 0 & 0 \end{bmatrix} \\
\mathbf{K}_4(s) =& \begin{bmatrix} 0 & 0 & 0 & (K_{pmx} + K_{imx} / s + K_{dmx} s) & 0 \end{bmatrix} \\
\mathbf{K}_5(s) =& \begin{bmatrix} 0 & 0 & 0 & 0 & (K_{imy} + K_{imy} / s + K_{dmy} s) \end{bmatrix}. \\
 \end{aligned}
\label{controller_mat}
\end{equation}
\normalsize
Taking the Laplace transformation of \eqref{plant3_2}, we obtain that
\begin{equation}
s \mathbf{X}(s) = \mathbf{A} \mathbf{X}(s) + \mathbf{B}_v \mathbf{V}(s).
\label{closeloop_stability1}
\end{equation}
Furthermore, $\mathbf{Y}(s)$ can be expressed as
\begin{equation}
\mathbf{Y}(s) = \mathbf{C} \mathbf{X}(s) + \mathbf{H} \mathbf{V}(s)
\label{closeloop_stability2}
\end{equation}
where $\mathbf{C}$ and $\mathbf{H}$ are given as
\footnotesize
\begin{equation}
\mathbf{C}^T=
\begin{bmatrix}
V_0 & 0 & 0 & 0 & 0 \\
\rho C_d A & 1/V_0 & 0 & 0 & 0 \\
0 & 0 & 1 & 0 & 0 \\
0 & 0 & 0 & 0 & 0 \\
0 & 0 & 0 & 0 & 0 \\
0 & 0 & 0 & 1 & 0 \\
0 & 0 & 0 & 0 & 0 \\
0 & 0 & 0 & 0 & 1 \\
0 & 0 & 0 & 0 & 0 \\
0 & 0 & 0 & 0 & 0 \\
0 & 0 & 0 & 0 & 0 \\
0 & 0 & 0 & 0 & 0 \\
0 & 0 & 0 & 0 & 0 \\
0 & 0 & 0 & 0 & 0 \\
0 & 0 & 0 & 0 & 0 \\
0 & 0 & 0 & 0 & 0 \\
0 & 0 & 0 & 0 & 0 \\
 \end{bmatrix}
\label{output_mat}
\end{equation}
\begin{equation}
\mathbf{H}=
\begin{bmatrix}
    1 & 0 & 0 & 0 & 0 \\
    0 & 0 & 0 & 0 & 0 \\
    0 & 0 & 0 & 0 & 0 \\
    0 & 0 & 0 & 0 & 0 \\
    0 & 0 & 0 & 0 & 0 \\
 \end{bmatrix}.
\label{ff_mat}
\end{equation}
\normalsize
Using \eqref{closeloop_stability1} and \eqref{closeloop_stability2}, we obtain that
\begin{equation}
\mathbf{Y}(s) = (\mathbf{C} (s \mathbf{I} - \mathbf{A})^{-1} \mathbf{B_v} + \mathbf{H} ) \mathbf{V}(s).
\label{closeloop_stability3}
\end{equation}
Defining $\mathbf{G}(s) \equiv    \mathbf{C} (s \mathbf{I} - \mathbf{A})^{-1} \mathbf{B_v} + \mathbf{H}$ and substituting \eqref{closeloop_stability0} into \eqref{closeloop_stability3}, the relationship between the reference signal and the output vector can be written as
\begin{equation}
\mathbf{Y}(s) = (s \mathbf{I} + \mathbf{G}(s) \mathbf{K}(s))^{-1} \mathbf{G}(s) \mathbf{K}(s) \  \mathbf{R}(s).
\label{closeloop_stability4}
\end{equation}
Considering the definitions,
\begin{equation}
\begin{aligned}
\mathbf{T_0}(s) =& (s \mathbf{I} + \mathbf{G}(s) \mathbf{K}(s))^{-1} \mathbf{G}(s), \mathbf{K}(s) \\
\mathbf{S_0}(s) =& (s \mathbf{I} + \mathbf{G}(s) \mathbf{K}(s))^{-1},
\end{aligned}
\label{closeloop_stability5}
\end{equation}
it can be shown that $\mathbf{T_0}(s)$, $\mathbf{S_0}(s)\mathbf{G}(s)$ and $\mathbf{K}(s)\mathbf{S_0}(s)$ are stable. Therefore the MIMO system, whose dynamics are given by \eqref{closeloop_stability0} - \eqref{closeloop_stability3} are internally stable  \cite{stability}.  

\end{document}